\documentclass[12pt]{article}

\usepackage{amsmath, amsfonts,dsfont,amssymb,rotating, booktabs, tikz, array, multirow, dcolumn}
\usepackage{graphicx,psfrag,epsf}
\usepackage{enumerate}
\usepackage{natbib}
\usepackage{url} 
\usepackage{hyperref}
\hypersetup{
    colorlinks=true,
    linkcolor=blue,
    citecolor=blue,
    filecolor=blue,      
    urlcolor=cyan,
}
\usepackage[toc,page]{appendix}
\usepackage[skip=3pt, font=small, labelfont=bf]{caption}

\newcommand{\blind}{0}

\graphicspath{{Images/}}

\newtheorem{defn}{Definition}
\newtheorem{assumpt}{Assumption}
\newtheorem{theorem}{Theorem}

\DeclareMathOperator{\E}{E}

\DeclareMathOperator{\Y}{Y}
\DeclareMathOperator{\W}{W}
\DeclareMathOperator{\w}{w}
\DeclareMathOperator{\X}{X}

\newcommand{\info}[2]{\mathcal{#1}_{#2}}
\newcommand{\MA}{MA}

\newcommand{\bs}[1]{\boldsymbol{#1}}

\begin{document}

\def\spacingset#1{\renewcommand{\baselinestretch}%
{#1}\small\normalsize} \spacingset{1}


\if0\blind
{
  \title{\bf Estimating the Causal Effect of an Intervention in a Time Series Setting: the C-ARIMA Approach}
  \author{Fiammetta Menchetti\footnote{Corresponding author, email \texttt{fiammetta.menchetti@unifi.it}} \hspace{.2cm}\\
    DiSIA, University of Florence\\
    and \\
    Fabrizio Cipollini \\
    DiSIA, University of Florence \\
    and \\
    Fabrizia Mealli \\
    DiSIA, University of Florence}
  \maketitle
} \fi

\if1\blind
{
  \bigskip
  \bigskip
  \bigskip
  \begin{center}
    {\LARGE\bf Estimating the Causal Effect of an Intervention in a Time Series Setting: the C-ARIMA Approach}
\end{center}
  \medskip
} \fi

\bigskip
\begin{abstract}
The Rubin Causal Model (RCM) is a framework that allows to define the causal effect of an intervention as a contrast of potential outcomes. In recent years, several methods have been developed under the RCM to estimate causal effects in time series settings. None of these makes use of ARIMA models, which are instead very common in the econometrics literature. In this paper, we propose a novel approach, C-ARIMA, to define and estimate the causal effect of an intervention in a time series setting under the RCM. We first formalize the assumptions enabling the definition, the estimation and the attribution of the effect to the intervention; we then check the validity of the proposed method with an extensive simulation study, comparing its performance against a standard intervention analysis approach. In the empirical application, we use C-ARIMA to assess the causal effect of a permanent price reduction on supermarket sales. The CausalArima R package provides an implementation of our proposed approach.
\end{abstract}

\noindent%
\vfill

\newpage
\spacingset{1.45} 

\section{Introduction}
\label{sect:intro}

The potential outcomes approach is a framework that allows to define the causal effect of a treatment (or “intervention”) as a contrast of potential outcomes, to discuss assumptions enabling to identify such causal effects from available data, as well as to develop methods for estimating  causal effects under these assumptions \citep{Rubin:1974,Rubin:1975,Rubin:1978,Imbens:Rubin:2015}. Following \citet{Holland:1986}, we refer to this framework as the Rubin Causal Model (RCM). Under the RCM, the causal effect of a treatment is defined as a contrast of potential outcomes, only one of which will be observed while the others will be missing and become counterfactuals once the treatment is assigned. For example, in a study investigating the impact of a new legislation to reduce carbon emissions, the incidence of lung cancer after the enforcement of the new law is the observed outcome, whereas the counterfactual outcome is the incidence that would have been observed if the legislation had not been enforced. 

Having its roots in the context of randomized experiments, several methods have been developed to define and estimate causal effects under the RCM in the most diverse settings, including networks \citep{VanderWeele:2010, Forastiere:Airoldi:Mealli:2020, Noirjean:Mariani:Mattei:Mealli:2020}, time series \citep{Robins:1986, Robins:Greenland:Hu:1999, Bojinov:Shephard:2019} and panel data \citep{Rambachan:Shephard:2019, Bojinov:Rambachan:Shephard:2020}. 

Focusing on a time series setting, a different approach extensively used in the econometric literature is \textit{intervention analysis}, introduced by \citet{Box:Tiao:1975,Box:Tiao:1976} to assess the impact of shocks occurring on a time series. Since then, it has been successfully applied to estimate the effect of interventions in many fields, including economics \citep{Larcker:Gordon:1980, Balke:Fomby:1994}, social science \citep{Bhattacharyya:Layton:1979,Murry:Stam:Lastovicka:1993} and terrorism \citep{Cauley:Iksoon:1988,Enders:Sandler:1993}. The effect is generally estimated by fitting an ARIMA-type model with the addition of an intervention component whose structure should capture the effect generated on the series (e.g., level shift, slope change and similar). However, this approach fails to define the causal estimands and to discuss the assumptions enabling the attribution of the uncovered effect to the intervention. 

Closing the gap between causal inference under the RCM and intervention analysis, in this paper we propose a novel approach, Causal-ARIMA (C-ARIMA), to estimate the causal effect of an intervention in observational time series settings where units receive a single persistent treatment over time. In particular, we introduce the assumptions needed to define and estimate the causal effect of an intervention under the RCM; we then define the causal estimands of interest and derive a methodology to perform inference. Laying its foundation in the potential outcomes framework, the proposed approach can be successfully used to estimate properly defined causal effects, whilst making use of ARIMA-type models that are widely employed in the intervention analysis literature.

We also test C-ARIMA with an extensive simulation study to explore its performance in uncovering causal effects as compared to a standard intervention analysis approach. The results indicate that C-ARIMA performs well when the true effect takes the form of a level shift; furthermore, it outperforms the standard approach in the estimation of irregular, time-varying effects. 

Finally, we illustrate how the proposed approach can be conveniently applied to solve real inferential issues by estimating the causal effect of a permanent price reduction on supermarket sales. More specifically, on October 4, 2018 the Florence branch of an Italian supermarket chain introduced a new price policy that permanently lowered the price of 707 store brands. The main goal is to assess whether the new policy has influenced the sales of those products. In addition, we want to estimate the indirect effect of the permanent price reduction on the perfect substitutes, i.e., products sharing the same characteristics of the discounted goods but selling under a competitor brand. Our results suggest that store brands' sales increased due to the permanent price discount; interestingly, we find little evidence of a detrimental effect on competitor brands, suggesting that unobserved factors may drive competitor-brand sales more than price. We therefore believe that our approach can support decision making at firms, as it can be used to perform business research whose implications can be of great interest to marketing professionals.  

The remainder of the paper is organized as follows: Section \ref{sect:lit} surveys the literature; Section \ref{sect:causal_framework} presents the causal framework; Section \ref{sect:cARIMA} illustrates the proposed C-ARIMA approach; Sections \ref{sect:simulations} and \ref{sect:data_method} describe, respectively, the simulations and the empirical study; Section \ref{sect:conclusions} concludes.

\section{Literature review}
\label{sect:lit}
In time series settings, the identification and the estimation of causal effects using potential outcomes have been formalized in the context of randomized experiments \citep{Bojinov:Shephard:2019, Rambachan:Shephard:2019, Bojinov:Rambachan:Shephard:2020}. However, unlike randomized experiments, in an observational study  the assignment mechanism, i.e., the process that determines which units receive treatment and which receive control, is unknown. Thus, observational studies pose additional challenges to the identification and estimation of causal effects, especially in time series settings, due to the presence of a single series receiving the intervention: estimands usually employed in panel settings like the ATT (average treatment effect on the treated units) are not applicable and sometimes it might be difficult to find suitable control series. 

A method that has been extensively used to evaluate the effect of interventions in the absence of experimental data is \textit{Difference-in-Difference} (DiD) \citep{Card:Krueger:1993, Meyer:Viscusi:Durbin:1995, Garvey:Hanka:1999, Angrist:Pischke:2008, Anger:Kvasnicka:Siedler:2011}. In its simplest formulation, this method requires to observe a treated and a control group at a single point in time before and after the intervention; the effect is then estimated by contrasting the change in the average outcome for the treated group with that of the control group under the assumption that, in the absence of treatment, the outcomes of the treated and control units would have followed parallel paths. However, many applications differ from this canonical setup: treatments may occur at different times, and the parallel trend assumption is often unrealistic \citep{Abadie:2005, Ryan:Burgess:Dimick:2015, ONeill:Kreif:Grieve:Sutton:Sekhon:2016}.\footnote{For example, in case of time-varying unobserved confounders the parallel trend assumption is invalid. In such cases, researchers may switch to methods relying on the assumption of ignorability conditional on past outcomes and covariates, like the \textit{lagged dependent variable estimator} (LDV). However, LDV and DiD estimators are related by a \textit{bracketing relationship} \citep{Angrist:Pischke:2008, Ding:Li:2019}, namely, if either parallel trends or ignorability holds, the true effect is bounded by the LDV and the DiD estimators, so when we rely on the wrong assumption we will either overestimate or underestimate the effect. In practice, when multiple pre-treatment periods are available, researchers test for parallel trends before employing DiD. \citet{Roth:2018} also suggests to adjust for the result of pretesting and \citet{Rambachan:Roth:2019} propose new methods that weakens the reliance on the parallel trend assumption by imposing restrictions on the differences in trends between treated and control units.}

Emerging literature on heterogeneous treatment effects in DiD with staggered adoption and variation in timing have partially overcome these limitations. For example, \citet{Callaway:Santanna:2020} introduce a weaker version of the parallel trend assumption that holds after conditioning on covariates; in addition, by focusing on a design-based perspective in which the assignment date is assumed to be randomized, \citet{Athey:Imbens:2021} do not need any functional form assumption for the potential outcomes. Furthermore, in a staggered adoption setting where the treatment time varies across units and they remain exposed to this treatment at all times afterwards, it is possible to estimate the effect of the intervention even when all units are eventually treated by using as a control group the set of last treated units \citep{Sun:Abraham:2020} or the set of not-yet treated units \citep{Callaway:Santanna:2020}. 
 
Another popular method to infer the causal effect of an intervention from panel data under the RCM is constructing a \textit{synthetic control} from a set of time series that are not directly impacted by the treatment and have weighted pre-treatment variables matching those of the treated unit \citep{Abadie:Gardeazabal:2003, Abadie:Diamond:Hainmueller:2010, Abadie:Diamond:Hainmueller:2015}. In contrast to DiD, synthetic control methods compensate for the lack of parallel trends by re-weighting control units so that the weighted pre-intervention outcomes and covariates are as close as possible to the average pre-intervention outcomes and covariates of the treated units. For example, in a study investigating the impact of a new legislation to reduce pollution levels, a suitable set of control series could be the evolution of carbon emissions in neighboring states that did not activate the new regulation: the synthetic control would then be constructed such that, in the pre-intervention period, the weighted average of the emissions and characteristics (e.g., population density, number of industries) of the neighboring states are similar to the emissions and characteristics of the treated state. Since their introduction, synthetic control methods have been successfully applied in a wide range of research areas, including healthcare \citep{Kreif:Grieve:Hangartner:Turner:Nikolova:Sutton:2016,Choirat:Dominici:Mealli:Papadogeorgou:Wasfy:Zigler:2018,Viviano:Bradic:2019}, economics \citep{Billmeier:Nannicini:2013,Abadie:Diamond:Hainmueller:2015,Dube:Zipperer:2015,Gobillon:Magnac:2016,Benmichael:Feller:Rothstein:2018}, marketing and online advertising \citep{Brodersen:Gallusser:Koehler:Remy:Scott:2015,Li:2019}. Recently, \citet{Arkhangelsky:Athey:Hirshberg:Imbens:Wager:2019} introduced the synthetic difference in difference estimator (SDID) combining the attractive features of both DiD and synthetic controls: like synthetic controls, this method re-weights control units such that their weighted pre-intervention trend and the (average) trend of the treated unit(s) are parallel (but not necessarily identical), thereby weakening the reliance on the parallel trend assumption; then, it uses DiD on the re-weigted panel by also focusing on the time periods that are more similar to the post-intervention periods. Therefore, SDID also addresses the concerns on pre-trend adjustments expressed in \citet{Roth:2018}. 

Nevertheless, DiD estimators, synthetic control methods and their combinations usually require to observe at least one suitable control unit, which is often impractical. For example, in our application, appropriate control series could be the sales of products that are not impacted by the new policy. However, since the supermarket chain implemented an extensive price policy change addressing at least one product in each category, all products were impacted directly or indirectly by the intervention, preventing us to find suitable controls. In addition, all products received the intervention simultaneously, thereby precluding the adoption of the DiD estimators developed under variation in timing. Finally, by focusing on a few time points, DiD estimators and synthetic controls have a limited ability to exploit the information provided by pre-treatment temporal dynamics; thus, they are not the best option when there are few units observed over a large period of time (small N, large T panels).

A recent approach overcoming these limitations is proposed by \citet{Brodersen:Gallusser:Koehler:Remy:Scott:2015}. Their methodology share several features with DiD and synthetic control methods but, instead of using control units or external characteristics, it only requires to learn the dynamics of the treated unit prior to the intervention. In other words, it builds a synthetic control by forecasting the counterfactual series in the absence of intervention based on a model estimated on the pre-intervention data. In particular, the authors employ Bayesian Structural Time Series models \citep{West:Harrison:2006, Harvey:1989} since they allow to add the components (e.g., trend, seasonality, cycle) that better describe the characteristics of the time series, whilst incorporating prior knowledge in the estimation process. Borrowing the name from the associated \texttt{R} package, from now on we refer to their method as ``CausalImpact''. An extension of this approach to a multivariate time series setting is proposed in \citet{Menchetti:Bojinov:2020}, where the authors employ Multivariate Bayesian Structural Time Series model to assess the impact of an intervention on statistical units showing interactions with one another. 

Our work is closely related to DiD with staggered adoption, to synthetic control methods and to the methodology proposed by \citet{Brodersen:Gallusser:Koehler:Remy:Scott:2015}. In the same vein as CausalImpact, we propose C-ARIMA as a novel approach to build a synthetic control for a time series subject to an intervention by learning its time dynamics in the pre-intervention period and then forecasting the series in the absence of intervention. Both CausalImpact and C-ARIMA brings several improvements over DiD estimators and synthetic control methods: they are tailored to estimate the effect of an intervention when no control series is available; and they are well suited to the case of a single time series or small N large T panels, since they allow to fully exploit useful information provided by the pre-intervention dynamics.\footnote{On the contrary, when there are multiple units observed over a short period of time, DiD estimators, synthetic control methods and SDID can be a better choice, since they allow to exploit the panel dimension; moreover, when there is variation in treatment timing it would be possible to construct estimators for the average treatment effect even in the absence of untreated units.} Furthermore, compared to CausalImpact, our methodology is based on ARIMA models and thus can be used as an alternative by researchers and practitioners in a wide range of fields that are not familiar to (or are not willing to adopt) Bayesian inference. In addition, ARIMA models are able to describe a wide variety of time series generated by complex, non-stationary processes and are already implemented in a large number of statistical software programs, which makes C-ARIMA very easy to use in practice. 

Finally, the C-ARIMA also shares many features with the approach described in \citet{Box:Tiao:1976}, where the authors suggest to compare the observed data after an intervention with the forecasts from a model fitted to the pre-intervention period. In particular, the $k$-step ahead forecast error in \citet{Box:Tiao:1976} is equivalent to our point causal effect $\tau_{t+k}$ and, as a result, our test statistic as defined by equation (\ref{eqn:arima_var_tau}) in Section \ref{subsect:estimators} relates to their Q test statistic. However, in addition to \citet{Box:Tiao:1976} we also provide test statistics for two additional effects: the cumulative and the temporal average effect. Indeed, oftentimes researchers are interested in a cumulative sum of point effects. For example, in a study evaluating the effect of the Hospital Readmission Reduction Program on hospital readmissions and mortality, \citet{Choirat:Dominici:Mealli:Papadogeorgou:Wasfy:Zigler:2018} focus on estimating the total number of additional readmissions due to the new law over the entire post-intervention period. Most importantly, we frame our estimators in the potential outcome framework, defining the effects and discussing the assumptions enabling their attribution to the intervention: both \citet{Box:Tiao:1976} and canonical intervention analysis (see \citet{Box:Tiao:1975} and \citet{Bhattacharyya:Layton:1979} among others) fail to do that and thus it not clear whether their effect might also have a causal interpretation or not.
The distinction between C-ARIMA and canonical intervention analysis is even more apparent: in those setups, researchers need to make an assumption on the structural form of the effect (e.g., level shift); then, the resulting ARIMA model is estimated to the entire time series and the assumed structure is finally checked by analysing model adequacy. Essentially, this is a trial and error process: if model adequacy is poor, it needs to be re-estimated under a different assumed effect structure (e.g., slope change). Conversely, with C-ARIMA we do not need to impose any structure on the effect of the intervention.

\section{Causal framework}
\label{sect:causal_framework}
On October 4, 2018 the Florence branch of an Italian supermarket chain introduced a new price policy that permanently lowered the price of $707$ store brands in several product categories; the empirical analysis focuses on the goods belonging to the ``cookies'' category. The supermarket chain also sells competitor brand cookies with the same characteristics (e.g., ingredients, flavour, shape) as their store brand equivalent; starting from the intervention date, these products became more expensive compared to the store brands. Therefore, we can consider all of them to be treated: the treatment on the store brand is the permanent price reduction, whereas the treatment on the competitor brands is the resulting relative price increase. The main goal is to assess the overall impact of the price policy, which is done by estimating the causal effect of the two treatments on the sales of store and competitor brand cookies.

In this section we present the notation and discuss some assumptions allowing the estimation of the causal effect and its attribution to the intervention. We also deduce some examples from our empirical context, so as to clarify both the theoretical concepts and the application. Finally, we define the causal estimands we are interested in.  

\subsection{Assumptions}
\label{subsect:assumptions}

Let $\W_{j,t} \in \{0,1\}$ be a random variable describing the treatment assignment of unit $j \in \{1,\dots,J\}$ at time $t \in \{1,\dots, T\}$, where $1$ denotes that a ``treatment'' (or ``intervention'') has taken place and $0$ denotes control. The first assumption is about the treatment process.

\begin{assumpt}[Single persistent intervention]
\label{assumpt:single_int}
We say that unit $j$ received a single persistent intervention if there exists e $t^*_j \in \{ 1, \dots, T\}$ such that $\W_{j,t} = 0$ for all $t \leq t^*_i$ and $\W_{j,t} = \W_{j,t'} = \W_j$ for all $t,t' > t^*_j$. If all units receive a single intervention, then we say the study is a single intervention panel study. If the intervention happens simultaneously on all
units, that is, $t^*_j = t^*_{j'} = t^*$ we say the study is a simultaneous intervention panel study.
\end{assumpt}

In a randomized experiment, the treatment can be administered at any point in time \citep{Bojinov:Shephard:2019}, whereas in observational studies it is not uncommon to observe a single persistent treatment, as in the case of a policy change \citep{Callaway:Santanna:2020}, a new government law \citep{Choirat:Dominici:Mealli:Papadogeorgou:Wasfy:Zigler:2018} or a price promotion \citep{Brodersen:Gallusser:Koehler:Remy:Scott:2015}. This is also the case of our empirical application and, as a result, we restrict our attention to a setting where all units are subject to a simultaneous and persistent intervention and we denote with $t^{*}$ the intervention date. Assumption \ref{assumpt:single_int} is equivalent to the \textit{irreversibility of treatment} assumption in \citet{Callaway:Santanna:2020}; such a persistent treatment is also analogous to the \textit{absorbing treatment} in \citet{Sun:Abraham:2020}.

Whether a unit is assigned to treatment or control may impact its outcome. For example, a brand of cookies will likely sell more under a $50\%$ price discount compared to a scenario where it is not discounted. Under the RCM, the sales in these two alternative scenarios are known as potential outcomes. 

Denote with $\W_{1:J,1:T} = (\W_{1,1:T}, \dots, \W_{J,1:T})$ the assignment paths of all units up to time $T$ and let $\w_{1:J,1:T}$ be a realization of $\W_{1:J,1:T}$. In general, the potential outcomes of a unit $j$ at time $t$ are function of the entire assignment panel, i.e. $\Y_{j,t}(\w_{1:J, 1:T})$ \citep{Bojinov:Rambachan:Shephard:2020}. However, under Assumption \ref{assumpt:single_int} we are able to restrict this dependence structure by focusing on non-anticipating potential outcomes.

\begin{assumpt}[Non-anticipating potential outcomes]
\label{assumpt:non-anticipating outcomes}
For all $j \in \{1, \dots, J \}$, the outcome of unit $j$ at time $t \in \{1,\dots,t^*\}$ is independent of the treatment occurring at time $t^*$
$$\Y_{j,t}(\w_{1:J, 1:T}) = \Y_{j, t}(\w_{1:J, 1:t^*}).$$

\end{assumpt}

In words, pre-intervention outcomes are not impacted by the future intervention; an implication is that there is no treatment effect in the pre-treatment period. Assumption \ref{assumpt:non-anticipating outcomes} is analogous to the non-anticipation assumptions usually made in the literature \citep{Bojinov:Shephard:2019, Callaway:Santanna:2020, Sun:Abraham:2020} and it is plausible when the statistical units have no knowledge of the future intervention. This is also the case of our empirical application, since the supermarket chain did not advertise the price reduction in advance. Furthermore, in our empirical setting, we can also rule out any form of interference between the units (i.e., cookies) in each group of store and competitor brands.

\begin{assumpt}[Temporal no-interference] 
\label{assumpt:no-interference} 
For all $j \in \{1,\dots,J \}$ and all $t \in \{t^*+1,\dots,T\}$, the outcome of unit $j$ at time $t$ depends solely on its own treatment path
$$\Y_{j,t}(\w_{1:J, t^*+1:t}) = \Y_{j,t}(\w_{j, t^*+1:t}).$$
\end{assumpt}

It means that whether the other units receive the treatment at time $t^*$ or not, this doesn't affect unit $j$'s potential outcomes. This assumption, which is also known as Temporal Stable Unit Treatment Value Assumption or TSUTVA \citep{Bojinov:Shephard:2019,Bojinov:Rambachan:Shephard:2020}, is the time series equivalent of the cross-sectional SUTVA \citep{Rubin:1974} and contributes to reduce the number of potential outcomes. In our empirical setting, the cookies selected for the permanent price discount differ on many characteristics, such as the shape, flavor and the ingredients, meaning that they appeal to different customers. Therefore, the temporal no-interference assumption is plausible within each group of store and competitor brands. \\

Although playing a partially different role, above assumptions are crucial for a proper definition of the causal effect: Assumptions \ref{assumpt:single_int} and \ref{assumpt:no-interference} pose a limit to the set of potential outcomes; Assumption \ref{assumpt:non-anticipating outcomes} excludes any treatment effect anticipation. Moreover, they allow us to ease notation. Indeed, a single persistent intervention implies $\w_{j,t} = \w_{j,t'} = \w_j$ for all $j$ and all $t,t' \in \{t^*+1, \dots, T \}$, whereas for all $t \in \{1, \dots, t^* \}$ we simply have $\w_j = 0$. Under temporal no-interference we can also drop the $j$ subscript, so from now on we use $\Y_{t}(\w)$ to indicate the \textit{potential outcomes time series} of a generic unit at time $t$. 

Recall that there are two possible treatments for each unit in the post intervention period: $\w = 1$ if the unit is treated; $\w = 0$ if it is assigned to control. However, we can only observe one of them and in our empirical application, the observed path is $\w = 1$ for all units. Therefore, the observed potential outcome time series for all $t\in \{t^*+1, \dots, T \}$ is $\Y_{t}(1)$, whereas $\Y_{t}(0)$ is the \textit{missing} or \textit{counterfactual} potential outcome time series and needs to be estimated in order to compute the causal effect of the intervention. Including covariates that are linked to the outcome can improve the estimation of the missing potential outcome; however, if the covariates are influenced by the treatment, the estimates will be biased. We therefore make the following assumption.

\begin{assumpt}[Covariates-treatment independence]
\label{assumpt:covariates}
Let $\X_{t}$ be a vector of covariates that are predictive of the outcome; for all $t \in \{t^*+1,\dots,T \}$ such covariates are not affected by the intervention
$$\X_{t}(1) = \X_{t}(0) . $$
\end{assumpt}

As a result, we can use the known covariates values at time $t \in \{t^*+1, \dots, T \}$ to improve the prediction of the outcome in the absence of intervention $\Y_{t}(0)$. As detailed in Section \ref{sect:data_method}, our set of covariates includes a holiday dummy, some day-of-the-week dummies and the price per unit. While it is quite obvious that all the dummies are unaffected by the intervention, things get trickier for price. For the analysis on competitor brands we used their actual price, since it is not directly affected by the intervention; conversely, for the analysis on store brands, to avoid dependencies between treatment and price we considered a ``modified'' price that, starting from the intervention date, assumes a constant value equal to the price of day before the intervention, which is the most likely price that the item would have had if the discount hadn't been introduced.\footnote{The supermarket chain sometimes run temporary promotions reducing the price of selected goods for a limited period of time. The time interval after the permanent price discount spans from October 4, 2018 to April 30, 2019 and in the corresponding period before the intervention (October 4, 2017 - April 30, 2018) there were no temporary promotions on the store brands that are part of this analysis. Thus, the assumption of a constant price level in the period following the intervention is plausible.} 

Our last assumption entails the assignment mechanism and it is essential to ensure that the uncovered causal effect can be attributed to the intervention.

\begin{assumpt}[Non-anticipating treatment]
\label{assumpt:non-anticipating treatment}
The assignment mechanism at time $t^*$ depends solely on past outcomes and past covariates
$$\Pr(\W_{t^*} = \w_{t^*}| \W_{1:t^*-1}, \Y_{1:T}(\w_{1:T}), \X_{1:T}) = \Pr(\W_{t^*} = \w_{t^*}| \Y_{1:t^*-1}(\w_{1:t^*-1}), \X_{1:t^*-1}).$$
\end{assumpt}

In words, the decision of lowering the price of a product is informed only by its past sales performances and past covariates or, at most, by general beliefs on the sales evolution under active treatment. Instead, the assumption would be violated, for example, if prices were lowered to discourage the opening of a competing supermarket chain store in the neighborhood: in this case we would be uncertain if a positive effect on sales was due to the price reduction or to the deferred market entrance of the competitor. This assumption can not be tested, but based on the statements made by the supermarket chain, it is plausible in our empirical setting. Notice that Assumption \ref{assumpt:non-anticipating treatment} is a re-statement of the non-anticipating treatment assumption of \citet{Bojinov:Shephard:2019} in a setting with a single intervention occurring at time $t^*$.  Furthermore, a non-anticipating treatment in a time series setting is the analogous of the unconfounded assignment mechanism in the cross-sectional setting \citep{Imbens:Rubin:2015}. Whilst a classical randomized experiment is unconfounded by design, we focus on observational studies where we have no control on the assignment mechanism. Thus, the non-anticipating treatment assumption is essential to ensure that any difference in the potential outcomes is due to the
treatment.

\subsection{Causal Estimands}
\label{subsect:estimands}

We now introduce three related causal estimands: the \textit{point} effect (an instantaneous effect at each point in time after the intervention), the \textit{cumulative} effect (a partial sum of the point effects) and the \textit{temporal average} effect (the average of the point effects in a given time period). 

\begin{defn}
\label{def:point effect}
At time $t \in \{t^*+1, \dots, T \}$, the point causal effect is,

\begin{equation}
\label{eqn:tau}
\tau_{t}(1;0) = \Y_{t}(1) - \Y_{t}(0) 
\end{equation}

the cumulative causal effect is
\begin{equation}
\label{eqn:cum_tau}
\Delta_{t}(1;0) = \sum\limits_{s = t^*+1}^{t} \tau_{s}(1;0) 
\end{equation}

\label{def:avg effect}
the temporal average causal effect is
\begin{equation}
\label{eqn:avg_tau}
\bar{\tau}_{t}(1;0) = \frac{1}{t-t^*} \sum\limits_{s=t^*+1}^{t} \tau_{s}(1;0) = \frac{\Delta_{t}(1;0)}{t-t^*} .
\end{equation}
\end{defn}

In words, the point effect measures the causal effect at a specific point in time and can be defined at every $t \in \{t^*+1, \dots, T\}$, thereby originating a vector of causal effects. The cumulative effect is then obtained by summing the point effects up to a predefined time point. For example, in our application, the cumulative effect would be the total number of cookies sold due to the permanent price reduction from the day when the new policy became effective until the end of the analysis period. Finally, we can also compute an average causal effect for the period. In our case, the temporal average effect indicates the number of cookies sold daily, on average, due to the permanent price reduction. 

Notice that the point effect is analogous to the general causal effect defined in \citet{Bojinov:Shephard:2019}, with the difference that our estimand is referred to a special setting were the units are subject to a single persistent treatment. 

In the next section, we introduce the C-ARIMA model and we describe how it can be used to estimate the causal quantities of interest.

\section{C-ARIMA}
\label{sect:cARIMA}

We propose a causal version of the widely used ARIMA model, which we indicate as C-ARIMA. We first introduce a simplified version of this model for stationary data generating processes; then, we relax the stationarity assumption and we extend the model to encompass seasonality and external regressors. Finally, we provide a theoretical comparison of the proposed approach with a standard ARIMA model with no causal connotation, REG-ARIMA henceforth.

\subsection{Simplified framework}
\label{sub:simplified}
We start with a simplified model for stationary data generating processes: this allows us to illustrate the building blocks of our approach with a clear and easy-to-follow notation. So, let us assume the potential outcome series $\{ \Y_t(\w)\}$ evolving as

\begin{equation}\\
\label{eqn:arima_simple_w}
\Y_t(\w)  = c + \frac{\theta_q(L) }{\phi_p(L)}\varepsilon_t + \tau_t \mathds{1}_{\{\w = 1\}} 
\end{equation}

where: $c$ is a constant term; $\phi_p(L)$ and $\theta_q(L)$ are lag polynomials with $\phi_p(L)$ having all roots outside the unit circle; $\varepsilon_t$ is white noise with mean $0$ and variance $\sigma^2_{\varepsilon}$; $\tau_t = 0$ $\forall t \leq t^*$ and $\mathds{1}_{\{\w = 1\}}$ is an indicator function which is one if $\w = 1$. As a result, $\tau_t$ can be interpreted as the point causal effect at time $t \in \{t^*+1, \dots, T \}$, since it is defined as a contrast of potential outcomes,

$$\tau_t(\w = 1; \w = 0) \equiv \Y_t(\w = 1) - \Y_t(\w = 0) = \tau_t.$$

Notice that under Assumptions \ref{assumpt:non-anticipating outcomes}-\ref{assumpt:non-anticipating treatment}, $\tau_t$ is a properly defined causal effect in the RCM and, as such, it should not be confused with additive outliers or any other kind of intervention component typically used in the econometric literature (e.g., \citet{Box:Tiao:1975}, \citet{Chen:Liu:1993}). Indeed, we can show that Equation (\ref{eqn:arima_simple_w}) encompasses all types of interventions. For example, consider the following model specification (innovation-type effect), 

$$\Y_t(\w)  = c + \frac{\theta_q(L) }{\phi_p(L)}(\varepsilon_t + \tau_t\mathds{1}_{\{\w = 1\}})$$  

and define $\tau_t^{\dagger} = \frac{\theta_q(L) }{\phi_p(L)} \tau_t$. Then, we have 

$$\Y_t(\w)  = c + \frac{\theta_q(L) }{\phi_p(L)}\varepsilon_t + \tau_t^{\dagger}\mathds{1}_{\{\w = 1\}}$$ 

where $\tau_t^{\dagger}(1;0) = \Y_t(1) - \Y_t(0) = \tau_t^{\dagger}$ is the point causal effect at time $t$. As it will be clear in Section \ref{subsect:estimators}, our model is estimated on the pre-intervention data, thus in the C-ARIMA approach we do not need to find the structure that better represents the effect of the intervention (e.g., additive outlier, transient change, innovation outlier). Conversely, such effect emerges as a contrast of potential outcomes in the post-intervention period and the proposed approach allows us to estimate $\tau_t$ whatever structure it has. 

To improve readability of the model equations, from now on we use $\Y_t$ to indicate $\Y_t(\w)$; the usual notation is resumed in Section \ref{subsect:estimators}. Thus, Equation (\ref{eqn:arima_simple_w}) can be written as,

\begin{equation}
\label{eqn:arima_simple}
\Y_t  = c + \frac{\theta_q(L) }{\phi_p(L)}\varepsilon_t + \tau_t .
\end{equation}

Setting

\begin{equation*}
z_t =  \frac{\theta_q(L) }{\phi_p(L)}\varepsilon_t ,
\end{equation*}

Equation (\ref{eqn:arima_simple}) becomes,

\begin{equation*}
		\Y_t = c + z_{t} + \tau_{t}.
\end{equation*}
 
Assuming perfect knowledge of the parameters ruling $\{z_{t}\}$, indicating with $\info{I}{t^*}$ the information up to time $t^*$ and denoting with $H_{0}$ the situation where the intervention has no effect (namely, 
$\tau_t = 0$ for all $t > t^*$)
we have that for a positive integer $k$, the $k$-step ahead forecast of $\Y_t$ under $H_{0}$ conditionally on $\info{I}{t^*}$ is

\begin{equation}
\hat{\Y}_{t^* + k} = \E \left[ \Y_{t^* + k} | \info{I}{t^*}, H_{0} \right] = c + \hat{z}_{t^* + k|t^*} 
\end{equation}

where $\hat{z}_{t^* + k|t^*} = E[z_{t^*+k} | \info{I}{t^*}, H_{0}]$. Thus, $\hat{z}_{t^* + k|t^*}$ represents an estimate of the missing potential outcomes in the absence of intervention, i.e., $\hat{\Y}_{t^*+k}(0) = \hat{z}_{t^* + k|t^*}$.

\subsection{General framework}
\label{sub:general}

We now generalize the above framework to a setting where $\{ \Y_t \}$ is non-stationary and possibly includes seasonality as well as external regressors. 

Let $\{ \Y_t \}$ follow a regression model with ARIMA errors and the addition of the point effect $\tau_t$,

\begin{equation}
\label{eqn:arima_general}
(1-L^s)^D (1-L)^d \Y_t  =  \frac{\Theta_Q (L^s)\theta_q(L) }{ \Phi_P(L^s) \phi_p(L)}\varepsilon_t +  (1-L^s)^D (1-L)^d \X_t' \beta + \tau_t
\end{equation}

where $\Theta_Q (L^s)$, $\Phi_P(L^s)$ are the lag polynomials of the seasonal part of the model with $\Phi_P(L^s)$ having roots all outside the unit circle; $\X_t$ is a set of external regressors satisfying Assumption \ref{assumpt:covariates}; $(1-L^s)^D$ and $(1-L)^d$ are contributions of the differencing operators to ensure stationarity, and $s$ is the seasonal period. Notice that the intercept defined in model (\ref{eqn:arima_simple}) is now included in the set of regressors. To ease notation, defining

\begin{equation}
\label{eqn:z}
z_t = \frac{\Theta_Q (L^s)\theta_q(L) }{ \Phi_P(L^s) \phi_p(L)}\varepsilon_t
\end{equation}
 
and indicating with $T(\cdot)$ the transformation of $\Y_t$ needed to achieve stationarity, i.e. $T(\Y_t) =(1-L^s)^D (1-L)^d \Y_t $, model (\ref{eqn:arima_general}) becomes

\begin{equation*}
S_t = T(\Y_t) - T(\X_{t})' \beta   = z_{t} + \tau_{t} ,
\end{equation*}

where $T(\X_t)' = (1-L^s)^D (1-L)^d \X_t'$ indicates that the same transformation is applied to the vector of regressors. Thus, the $k$-step ahead forecast of $S_t$ under $H_0$, given the information up to time $t^*$ is 

$$\hat{S}_{t^*+k} = \E[S_{t^*+k} | \info{I}{t^*}, H_0] = \E[T(\Y_{t^*+k}) - T(\X_{t^*+k})' \beta | \info{I}{t^*}, H_0] = \hat{z}_{t^*+k|t^*} .$$ 

\subsection{Causal effect inference}
\label{subsect:estimators}
We now derive estimators for the causal effects defined in Section \ref{subsect:estimands} based on the C-ARIMA model and we discuss their properties.

\begin{defn}[Causal effect estimators] For any integer $k$, let $S_{t^*+k}(1)$ be the observed potential outcome time series and let $\hat{S}_{t^*+k}(0)$ be the corresponding estimate of the missing potential outcomes under model (\ref{eqn:arima_general}). Then, estimators of the point, cumulative and temporal average effects are, respectively,
\begin{equation}
  \label{eqn:tauhat}
  \hat{\tau}_{t^*+k}(1;0) = S_{t^*+k}(1) - \hat{S}_{t^*+k}(0)
\end{equation}
\begin{equation}
  \label{eqn:Deltahat}
  \hat{\Delta}_{t^*+k}(1;0) = \sum\limits_{h = 1}^{k} \hat{\tau}_{t^*+h}(1;0)
\end{equation}
\begin{equation}
  \label{eqn:avetauhat}
  \hat{\bar{\tau}}_{t^*+k}(1;0) = \frac{1}{k} \sum\limits_{h = 1}^{k} \hat{\tau}_{t^*+h}(1;0) = \frac{\hat{\Delta}_{t^*+k}(1;0)}{k}.
\end{equation}
\end{defn}

Inference on the point, cumulative and temporal average effects can be conducted using hypothesis tests based on the above estimators. The following theorem illustrates their distributional properties.

\begin{theorem}
\label{theo}
Let $\{\Y_t(\w)\}$ follow the regression model with ARIMA errors defined in Equation (\ref{eqn:arima_general}) and, for any $k > 0$, let 
$$H_0: \tau_{t^*+k}(1;0) = 0 $$ 
the null hypothesis that the intervention has no effect.
Then, the estimators of the point, cumulative and temporal average effects under $H_0$ can be expressed as
\begin{equation}
\label{eqn:tauhat-H0}
  \hat{\tau}_{t^*+k}(1;0) | H_0 = \sum_{i = 0}^{k-1} \psi_{i} \varepsilon_{t^* + k - i}
\end{equation}
\begin{equation}
\label{eqn:Deltahat-H0}
  \hat{\Delta}_{t^*+k}(1;0) 
  |H_0 = 
  \sum_{h = 1}^k \varepsilon_{t^*+h} \sum_{i = 0}^{k-h} \psi_i
\end{equation}
\begin{equation}
  \label{eqn:avetauhat-H0}
  \hat{\bar{\tau}}_{t^*+k}(1;0)  
  |H_0 = 
  \frac{1}{k} \sum_{h = 1}^k \varepsilon_{t^*+h} \sum_{i = 0}^{k-h} \psi_i,
\end{equation}
where the $\psi_i$'s are the coefficients of a moving average of order $k-1$ whose values are functions of the ARMA parameters in Equation (\ref{eqn:arima_general}). \\ 
In case the $\varepsilon_{t}$ error term is assumed normally distributed, Equations~(\ref{eqn:tauhat-H0})--(\ref{eqn:avetauhat-H0}) become
\begin{equation}
\label{eqn:arima_var_tau}
\hat{\tau}_{t^*+k}(1;0)|H_0 \sim N \left[ 
0 , 
\sigma^{2}_{\varepsilon} \sum_{i = 0}^{k - 1} \psi_{i}^{2}
\right]
\end{equation}

\begin{equation}
\label{eqn:arima_var_cumtau}
\widehat{\Delta}_{t^*+k}(1;0)|H_0 \sim N \left[ 
0,
\sigma^{2}_{\varepsilon} \sum_{h = 1}^{k} \left( \sum_{i = 0}^{k - h} \psi_{i} \right)^{2}	
\right]
\end{equation}

\begin{equation}
\label{eqn:arima_var_avgtau}
\hat{\bar{\tau}}_{t^*+k}(1;0)|H_0 \sim N \left[ 
0, 
\frac{1}{k^2} \sigma^{2}_{\varepsilon} \sum_{h = 1}^{k} \left( \sum_{i = 0}^{k - h} \psi_{i} \right)^{2}	
\right]
\end{equation}

Proof: given in Appendix \ref{appB_proof}.

\end{theorem}

Theorem~\ref{theo} implies that the testing procedure can be lead in two different ways.
If one safely relies on the normality of the error term, the test can be based on Equations~(\ref{eqn:arima_var_tau})--(\ref{eqn:arima_var_avgtau}) and the corresponding Gaussian quantiles.
Otherwise, one can compute empirical critical values from Equations~(\ref{eqn:tauhat-H0})--(\ref{eqn:avetauhat-H0}) by bootstrapping the errors from the model residuals.
In both cases, the needed parameters are replaced by their estimated counterpart. 
The implicit assumption is that the model can be misspecified in the real practice, but the possible misspecification is not so severe to jeopardize the consistency of the estimators of the model parameters and, in turn, of the $\psi_i$ estimators.  

So far, we derived estimators for three causal effects defined for the transformed variable $S_t = T(\Y_t)- T(\X_t)'\beta$, where $T(\cdot) = (1-L^s)^D (1-L)^d$ is a transformation to achieve stationarity. By doing a further step we can also estimate the effect for the original (untransformed) variable $\Y_t$. Indeed, model (\ref{eqn:arima_general}) can also be written as,

\begin{equation}
\label{eqn:arima_general_2}
\Y_t(\w)  =  \frac{\Theta_Q (L^s)\theta_q(L) }{(1-L)^d (1-L^s)^D \Phi_P(L^s) \phi_p(L)}\varepsilon_t +  \X_t' \beta + \tau^Y_t \mathds{1}_{\{\w = 1\}} 
\end{equation}

where 

\begin{equation}
    \label{eqn:tau_Y}
    \tau^Y_t = \frac{\tau_t}{(1-L)^d (1-L^s)^D}
\end{equation}

is the point causal effect on the original variable, defined by a contrast of two potential outcomes,

$$\tau^Y_t(\w = 1; \w = 0) \equiv \Y_t(\w = 1) - \Y_t(\w = 0).$$

Again, we can define the estimators for the point, cumulative and temporal average effects on the original variable under the null hypothesis that the intervention has no effect, $H_0: \tau^Y_t (1;0) = 0$ for all $t \in \{t^*+1, \dots, T \}$. 

\begin{defn}[Causal effect estimators on the original variable] For any integer $k$, let $\Y_{t^*+k}(1)$ be the observed, untransformed time series and let $\hat{\Y}_{t^*+k}(0)$ be the corresponding estimate of the missing potential outcomes under model (\ref{eqn:arima_general_2}). Then, estimators of the point, cumulative and temporal average effects on the original variable are, respectively,
\begin{equation}
  \label{eqn:tauhat_2}
  \hat{\tau}^Y_{t^*+k}(1;0) = \Y_{t^*+k}(1) - \hat{\Y}_{t^*+k}(0)
\end{equation}
\begin{equation}
  \label{eqn:Deltahat_2}
  \hat{\Delta}^Y_{t^*+k}(1;0) = \sum\limits_{h = 1}^{k} \hat{\tau}^Y_{t^*+h}(1;0)
\end{equation}
\begin{equation}
  \label{eqn:avetauhat_2}
  \hat{\bar{\tau}}^Y_{t^*+k}(1;0) = \frac{1}{k} \sum\limits_{h = 1}^{k} \hat{\tau}^Y_{t^*+h}(1;0) = \frac{\hat{\Delta}^Y_{t^*+k}(1;0)}{k}.
\end{equation}
\end{defn}

To perform inference on (\ref{eqn:tau_Y}) we introduce Theorem \ref{theo_2}, whose derivation follows directly from Theorem \ref{theo}.

\begin{theorem}
\label{theo_2}
Let $\{ \Y_t(\w) \}$ follow the regression model with ARIMA errors defined in Equation (\ref{eqn:arima_general_2}) and, for any $k > 0$, let $$H_0: \tau^Y_{t^*+k}(1; 0) = 0 $$ the null hypothesis that the intervention has no effect. Then, for some $b_i$ coefficients with $b_0 = 1$, the estimators of the causal effects on the original variable under $H_0$ can be expressed as,

\begin{equation}
\label{eqn:tauhat-H0_2}
  \hat{\tau}_{t^*+k}^{Y}(1;0) | H_0 = \sum_{j = 1}^{k} \varepsilon_{t^*+j} \sum\limits_{i = 0}^{k-j} b_i \psi_{k-j-i}
\end{equation}

\begin{equation}
\label{eqn:Deltahat-H0_2}
    \hat{\Delta}_{t^*+k}^{Y}(1;0) | H_0 = \sum_{h = 1}^k \varepsilon_{t^*+h} \sum_{i = 0}^{k-h} b_i \sum_{j = i}^{k-h} \psi_{k-h-j}
\end{equation}

\begin{equation}
\label{eqn:avetauhat-H0_2}
    \hat{\bar{\tau}}_{t^*+k}^{Y}(1;0) | H_0 = \frac{1}{k}\sum_{h = 1}^k \varepsilon_{t^*+h} \sum_{i = 0}^{k-h} b_i \sum_{j = i}^{k-h} \psi_{k-h-j}.
\end{equation}

where the $\psi_i$'s are the coefficients of a moving average of order $k-1$ whose values are function of the ARMA parameters in Equation (\ref{eqn:arima_general_2}). In case the $\varepsilon_t$ error term is normally distributed, Equations (\ref{eqn:tauhat-H0_2})--(\ref{eqn:avetauhat-H0_2}) become,

\begin{equation}
\label{eqn:arima_var_tau_2}
  \hat{\tau}_{t^*+k}^{Y}(1;0) | H_0 \sim N \left[ 0, \sigma^2_{\varepsilon} \sum_{j=1}^k \left( \sum_{i = 0}^{k-j} b_i \psi_{k-j-i} \right)^2 \right]
\end{equation}

\begin{equation}
\label{eqn:arima_var_cumtau_2}
    \hat{\Delta}_{t^*+k}^{Y}(1;0) | H_0 \sim N \left[ 0, \sum_{h = 1}^k \sigma^2_{\varepsilon} \left( \sum_{i = 0}^{k-h} b_i \sum_{j = i}^{k-h} \psi_{k-h-j}\right)^2 \right]
\end{equation}

\begin{equation}
\label{eqn:arima_var_avgtau_2}
    \hat{\bar{\tau}}_{t^*+k}^{Y}(1;0) | H_0 \sim N \left[ 0, \frac{\sigma^2_{\varepsilon}}{k^2} \sum_{h = 1}^k \left( \sum_{i = 0}^{k-h} b_i \sum_{j = i}^{k-h} \psi_{k-h-j}\right)^2 \right].
\end{equation}

Proof: given in Appendix \ref{appB_untransformed}.
\end{theorem}

\vspace{1em}

Summarizing, in order to estimate the causal effects (\ref{eqn:tau}), (\ref{eqn:cum_tau}) and (\ref{eqn:avg_tau}) we need to follow a three-step process: i) estimate the ARIMA model only in the pre-intervention period, so as to learn the dynamics of the dependent variable and the links with the covariates without being influenced by the treatment; ii) based on the process learned in the pre-intervention period, perform a prediction step and obtain an estimate of the counterfactual outcome during the post-intervention period in the absence of intervention; iii) by comparing the observations with the corresponding forecasts at any time point after the intervention, evaluate the resulting differences, which represent the estimated point causal effects. To perform inference, we can then use the results presented in Theorem \ref{theo} and Theorem \ref{theo_2}.

Conversely, REG-ARIMA model is fitted to the full time series (pre- and post-intervention) and the estimated coefficient for the dummy variable $D_t$ gives a measure of the association between the intervention (in the form of a level shift) and the outcome. 

\subsection{Comparison with REG-ARIMA}
To measure the effect of an intervention on an outcome repeatedly observed over time, a widely used approach is fitting a linear regression with ARIMA errors (REG-ARIMA). This method uses the entire time series and a dummy variable activating after the intervention; then, SARIMA-type errors are added to the model to account for autocorrelation and possible seasonality. In its simplest formulation, such a model can be written as,

\begin{align*}
\Y_t & = c + D_t \beta_0 + z_t \\
z_t & = \frac{\theta_q(L)}{\phi_p(L)} \varepsilon_t
\end{align*}

where $z_t$ is a stationary ARMA$(p,q)$; $D_t$ is a dummy variable taking value $1$ after the intervention and $0$ otherwise and $\beta_0$ is the regression coefficient. Generalizing to a possibly seasonal and non-stationary ARIMA, above model can be re-written as,

\begin{align}
\label{eqn:ARIMA_errors}
\Y_t &= \X_t' \beta + z_t \\ \nonumber
z_t &= \frac{\theta_q(L) \Theta_Q(L^s)}{(1-L)^d (1-L^s)^D \phi_p(L) \Phi_P(L^s)} \varepsilon_t
\end{align}

where $\X_t$ is a set of regressors, including the intercept and the dummy variable and $\beta$ is a vector of regression coefficients.\footnote{Notice that under REG-ARIMA, $\Y_t$ is not a potential outcome, therefore, it does not depend on the treatment path.} 

Essentially, REG-ARIMA is a standard intervention analysis approach that is used when the intervention is supposed to have produced a level shift on the outcome, i.e. a fixed change in the level of the outcome during the post-intervention period. Thus, there are two main differences between C-ARIMA and REG-ARIMA. First, without a critical discussion of the fundamental assumptions, the effect grasped by $\beta_0$ can not be attributed with certainty to the intervention. For example, it might be driven by an undetected confounder, biased by the inclusion of a regressor linked to the treatment, or even be the anticipated result of a future intervention. Second, the size of the effect is given by the estimated coefficient of a dummy variable activating after the intervention, so that REG-ARIMA can only capture effects in the form of level shifts. 
Conversely, C-ARIMA assumes no structure on $\tau_t$ and, as such, it can capture any form of effects (level shift, slope change and even irregular time-varying effects). Furthermore, the estimation of the effect is done in a very natural way under C-ARIMA. Indeed, intervention analysis requires the estimation of two models: the first learns the structure of the effect and the second measures its size; by letting the intervention component free to vary, C-ARIMA can instead estimate any form of effect in only one step.  

In Section \ref{sect:simulations} we report a simulation study where we compare the empirical performance of both approaches (C-ARIMA and REG-ARIMA) in inferring causal effects. 

\section{Simulation study}
\label{sect:simulations}
We perform a simulation study to check the ability of the C-ARIMA approach to uncover causal effects. Furthermore, in order to show its merits over a more standard approach, we also assess the performance of REG-ARIMA. We remark, however, that the comparison is purely methodological, since the theoretical limitations of REG-ARIMA do not allow the attribution of such effects to the intervention. Sections \ref{subsect:sim_design} and \ref{subsect:sim_res} illustrate, respectively, the simulations design and the results.   

\subsection{Design}
\label{subsect:sim_design}
We generate $1000$ replications from the following ARIMA$(1,0,1)(1,0,1)_7$ model,\footnote{Notice that $D=d=0$ implies $\tau_t \equiv \tau^Y_t$.}

\begin{align*}
\Y_t & = \beta_1 \X_{1,t} + \beta_2 \X_{2,t} + z_t \\
z_t & = \frac{\theta_q(L) \Theta_Q (L^s)}{\phi_p(L) \Phi_P (L^s)} \varepsilon_t .
\end{align*}

\begin{figure}[h!]
\centering
\caption{Evolution of the generated covariates, $\X_{1,t}$ and $\X_{2,t}$ and their combination according to the simulated model, $\beta_1 \X_{1,t}+\beta_2 \X_{2,t}$.}
\label{fig:sim_cov_plot}
\includegraphics[scale=0.4]{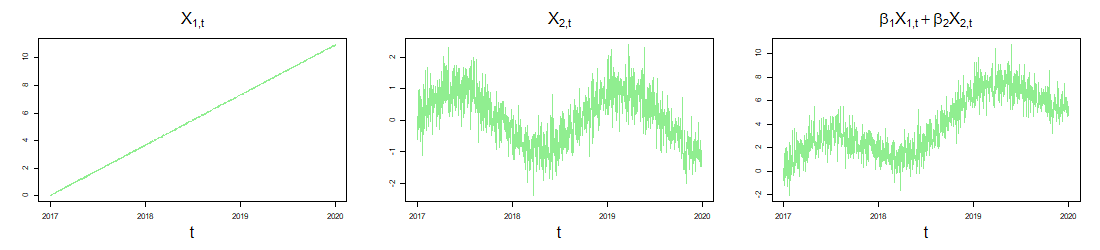}
\end{figure}

The two covariates of the regression equation are generated as $\X_{1,t} = \alpha_1t + u_{1,t}$ and $\X_{2,t} = \sin (\alpha_2t) + u_{2,t}$, with $\alpha_1 = \alpha_2 = 0.01$, $u_{1,t} \sim N(0,0.02)$, $u_{2,t} \sim N(0,0.5)$ and coefficients $\beta_1 = 0.7$ and $\beta_2 = 2$, respectively; regarding the ARIMA parameters, they are set to $\phi_1 = 0.7$, $\Phi_1 = 0.6$, $\theta_1 = 0.6$ and $\Theta_1 = 0.5$. Finally, $\varepsilon_t \sim N(0, \sigma)$ with $\sigma = 5$. Figure \ref{fig:sim_cov_plot} shows the evolution of the generated covariates and their linear combination according to the above model.

\begin{figure}[h!]
\centering
\caption{For the same simulated time series (denoted as ``control'') the plots display two different types of effect: on the left, a level shift of $25\%$; on the right, the time series under treatment follows an irregular pattern.}
\label{fig:sim_plot_1}
\includegraphics[scale=0.4]{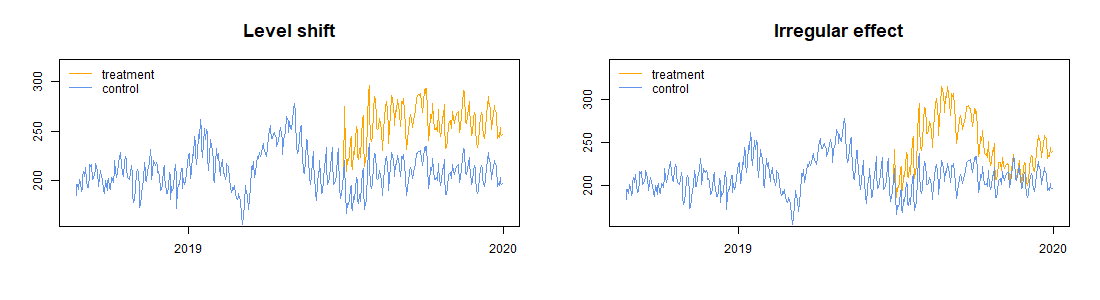}
\end{figure}

We assume that each generated time series starts on January 1, 2017 and ends on December 31, 2019 and that a fictional intervention takes place on June 30, 2019. In particular, we tested two types of intervention: i) a level shift with $5$ different magnitudes, i.e., $+1 \%$, $+10 \%$, $+25 \%$, $+50 \%$, $+100 \%$ ; ii) an intervention producing an immediate shock of $+10\%$ followed by a steady increase up to $+40\%$, a regular decline afterwards and a second increase near the end of the analysis period. As an example, Figure \ref{fig:sim_plot_1} provides a graphical representation of the two interventions for one of the simulated time series. 

The estimation of the causal effect is performed under two different models: the proposed C-ARIMA approach and REG-ARIMA, i.e, a linear regression with ARIMA errors and the addition of a dummy variable, as in Equation (\ref{eqn:ARIMA_errors}). Recall from previous Section \ref{subsect:estimators} that the C-ARIMA approach requires that the model is estimated on the pre-intervention data and the effect is given by direct comparison of the observed series and the corresponding forecasts post-intervention. Conversely, REG-ARIMA is fitted on the full time series and the estimated coefficient of the dummy variable provides a measure of the impact of the intervention. In addition, we estimate two versions of each model: a correctly specified model, denoted respectively with C-ARIMA$^{TRUE}$ and REG-ARIMA$^{TRUE}$, and the best fitting model selected by BIC minimization, denoted with C-ARIMA$^{BIC}$ and REG-ARIMA$^{BIC}$. Finally, in order to evaluate the performance of both approaches in uncovering causal effects at longer time horizons, we perform predictions at $1$ month, $3$ months and $6$ months from the intervention. As a result, the total number of estimated models in the pre-intervention period is $4000$ and the total number of estimated causal effects is $72,000$ (one for each time series, model, tested intervention and time horizon).

We measure the performance of the four models in terms of the following indicators: 

\begin{enumerate}
\item the length of the confidence interval around the true temporal average effect $\bar{\tau}_t$ for C-ARIMA$^{TRUE}$ and C-ARIMA$^{BIC}$ and around $\beta_0$ for REG-ARIMA$^{TRUE}$ and REG-ARIMA$^{BIC}$;
\item the absolute percentage error, defined for each model and intervention type as,
$$a_{i,h} = \frac{|\hat{\bar{\tau}}_{i,h} - \bar{\tau}_{i,h}|}{\bar{\tau}_{i,h}} \hspace{5pt} ; \hspace{5pt} a_{i,h}' = \frac{|\hat{\beta}_{0,i,h} - \bar{\tau}_{i,h}|}{\bar{\tau}_{i,h}} \hspace{10pt} i = 1,\dots, 1000 ; $$
where: $\hat{\bar{\tau}}_{i,h}$ and $\hat{\beta}_{0,i,h}$ indicate, respectively, the estimated causal effect and the estimated coefficient of the intervention dummy for the $i$-th simulated time series at time horizon $h$, where $h = \{1,2,3\}$ indicate the months after the intervention; $\bar{\tau}_{i,h}$ denotes the true temporal average effect (always positive);
\item the interval coverage, computed for the REG-ARIMA as the proportion of true effects in the estimated $95 \%$ confidence intervals over the $1000$ simulated series, whereas for the C-ARIMA it is obtained as the proportion of the true point effects within the estimated confidence intervals, which is then averaged over the simulated series.
\end{enumerate}

\subsection{Results}
\label{subsect:sim_res}

Table \ref{tab:sim_res_ci} shows the simulation results in terms of the length of the $95 \%$ confidence intervals around $\tau_t(1;0)$ and $\beta_0$, respectively. As expected, for the C-ARIMA models the interval length is independent of the impact; we can also notice that it reduces as the time horizon increases, whereas the interval length estimated under REG-ARIMA is stable over time. Finally, we can observe that REG-ARIMA yields shorter confidence intervals than C-ARIMA. 

Table \ref{tab:sim_res_err} reports the absolute percentage errors resulting from the simulations. When the intervention takes the form of a level shift, the error decreases with the size of the effect and, unsurprisingly, REG-ARIMA yields slightly better results than C-ARIMA. Indeed, the former model is especially suited for interventions in the form of level shifts. However, when we consider a level shift $> +50\%$ or an irregular intervention, the estimation errors of REG-ARIMA are $2$ to $4$ times higher than those coming from C-ARIMA. 

The interval coverage is reported in Table \ref{tab:sim_res_coverage}. Again, the coverage of the C-ARIMA approach does not change with the impact size and it is very close to the nominal $95 \%$ level. Instead, the coverage of REG-ARIMA decreases with the impact size and, with the only exception of the first two impacts, the results are quite far from the nominal $95\%$ level. This can be explained by the short confidence intervals achieved by REG-ARIMA, suggesting that even though the estimation error is small, the confidence intervals are not wide enough to contain the true effect. More importantly, when the effect is irregular, the estimated confidence intervals never contain the true effect.

Concluding, REG-ARIMA approach fails to detect irregular interventions and, most of the times, it does not achieve the desired interval coverage. As expected, REG-ARIMA model is suited only when there is reason to believe that the intervention produced a fixed change in the outcome level. Otherwise, should the researcher fail to identify the structure of the effect, using REG-ARIMA on irregular patterns produces biased estimates. Conversely, the C-ARIMA approach does a reasonably good job in detecting both type of interventions. Moreover, C-ARIMA does not require an investigation of the effect type prior to the estimation step; in addition, when the intervention is in the form of a level shift, the reliability of the C-ARIMA estimates increases with the impact size. Finally, we can observe that the results of the models selected through BIC minimization are very similar to those of the correct model specifications (the BIC correctly identifies the true model $74\%$ of times). 

\begin{table}[h!]
\centering
\caption{Length of the $95 \%$ confidence intervals around the estimated intervention effect $\tau_t(1;0)$ (for C-ARIMA) and $\beta_0$ (for REG-ARIMA). The different impact sizes ranging from $+1\%$ to $+100\%$ in the rows denote estimated effects in the form of level shifts, whereas NS stands for ``no structure'', thereby indicating the irregular effect. For each generated time series, impact size and time horizon ($1$, $3$ and $6$ months), the estimates are performed under two model specifications: the true model and the best fitting model based on BIC, denoted, respectively,  with the superscripts $TRUE$ and $BIC$.}
\label{tab:sim_res_ci}
\scalebox{0.8}{
\begin{tabular}{@{}lrrrrrrr@{}}
\toprule
     & \multicolumn{3}{c}{C-ARIMA$^{BIC}$}  &  & \multicolumn{3}{c}{REG-ARIMA$^{BIC}$}  \\ \cmidrule{2-4} \cmidrule{6-8}
$\tau_t(1;0)$    & 1 month   & 3 months  & 6 months &  & 1 month   & 3 months & 6 months \\ \midrule
$+1\%$   & 42.029 & 34.479 & 26.330 & & 10.620 & 10.458 & 10.340 \\ 
$+10\%$  & 42.029 & 34.479 & 26.330 & & 10.656 & 10.555 & 10.511 \\ 
$+25\%$  & 42.029 & 34.479 & 26.330 & & 10.734 & 10.751 & 10.849 \\ 
$+50\%$  & 42.029 & 34.479 & 26.330 & & 10.916 & 11.164 & 11.528 \\ 
$+100\%$ & 42.029 & 34.479 & 26.330 & & 11.454 & 12.259 & 13.233 \\ 
NS       & 42.027 & 34.474 & 26.325 & & 10.709 & 10.931 & 10.891 \\ 
\midrule
     & \multicolumn{3}{c}{C-ARIMA$^{TRUE}$} &  & \multicolumn{3}{c}{REG-ARIMA$^{TRUE}$} \\ \cmidrule{2-4} \cmidrule{6-8}
$\tau_t(1;0)$     & 1 month   & 3 months  & 6 months &  & 1 month   & 3 months & 6 months \\ \midrule
$+1\%$   & 42.055 & 34.536 & 26.381 & & 10.604 & 10.451 & 10.334 \\ 
$+10\%$  & 42.055 & 34.536 & 26.381 & & 10.639 & 10.548 & 10.506 \\ 
$+25\%$  & 42.055 & 34.536 & 26.381 & & 10.716 & 10.743 & 10.842 \\ 
$+50\%$  & 42.055 & 34.536 & 26.381 & & 10.895 & 11.152 & 11.520 \\ 
$+100\%$ & 42.055 & 34.536 & 26.381 & & 11.424 & 12.240 & 13.225 \\ 
NS       & 42.058 & 34.533 & 26.377 & & 10.691 & 10.913 & 10.876 \\
\bottomrule
\end{tabular}

}
\end{table}

\begin{table}[h!]
\centering
\caption{Absolute percentage error for the estimated intervention effect $\tau_t(1;0)$ (for C-ARIMA) and $\beta_0$ (for REG-ARIMA). The different impact sizes ranging from $+1\%$ to $+100\%$ in the rows denote estimated effects in the form of level shifts, whereas NS stands for ``no structure'', thereby indicating the irregular effect. For each generated time series, impact size and time horizon ($1$, $3$ and $6$ months), the estimates are performed under two model specifications: the true model and the best fitting model based on BIC, denoted, respectively,  with the superscripts $TRUE$ and $BIC$.}
\label{tab:sim_res_err}
\scalebox{0.8}{
\begin{tabular}{@{}lrrrrrrr@{}}
\toprule
     & \multicolumn{3}{c}{C-ARIMA$^{BIC}$}  &  & \multicolumn{3}{c}{REG-ARIMA$^{BIC}$}  \\ \cmidrule{2-4} \cmidrule{6-8}
 $\tau_t(1;0)$    & 1 month   & 3 months  & 6 months &  & 1 month   & 3 months & 6 months \\ \midrule
$+1\%$   & 4.185 & 3.783 & 3.405 & & 0.970 & 0.955 & 0.950 \\ 
$+10\%$  & 0.418 & 0.378 & 0.340 & & 0.116 & 0.118 & 0.117 \\ 
$+25\%$  & 0.167 & 0.151 & 0.136 & & 0.074 & 0.080 & 0.078 \\ 
$+50\%$  & 0.084 & 0.076 & 0.068 & & 0.065 & 0.072 & 0.071 \\ 
$+100\%$ & 0.042 & 0.038 & 0.034 & & 0.062 & 0.070 & 0.069 \\ 
NS       & 0.237 & 0.138 & 0.173 & & 0.423 & 0.610 & 0.463 \\ 
\midrule
     & \multicolumn{3}{c}{C-ARIMA$^{TRUE}$} &  & \multicolumn{3}{c}{REG-ARIMA$^{TRUE}$} \\ \cmidrule{2-4} \cmidrule{6-8}
$\tau_t(1;0)$     & 1 month   & 3 months  & 6 months &  & 1 month   & 3 months & 6 months \\ \midrule
$+1\%$   & 4.182 & 3.775 & 3.398 & & 0.963 & 0.946 & 0.943 \\ 
$+10\%$  & 0.418 & 0.378 & 0.340 & & 0.115 & 0.118 & 0.116 \\ 
$+25\%$  & 0.167 & 0.151 & 0.136 & & 0.074 & 0.079 & 0.078 \\ 
$+50\%$  & 0.084 & 0.076 & 0.068 & & 0.065 & 0.072 & 0.071 \\ 
$+100\%$ & 0.042 & 0.038 & 0.034 & & 0.062 & 0.070 & 0.068 \\ 
NS       & 0.237 & 0.137 & 0.172 & & 0.424 & 0.610 & 0.463 \\ 
\bottomrule
\end{tabular}

}
\end{table}

\begin{table}[h!]
\centering
\caption{Interval coverage in percentage of the true effects within the estimated intervals around $\tau_t(1;0)$ (for C-ARIMA) and $\beta_0$ (for REG-ARIMA). The different impact sizes ranging from $+1\%$ to $+100\%$ in the rows denote estimated effects in the form of level shifts, whereas NS stands for ``no structure'', thereby indicating the irregular effect. For each generated time series, impact size and time horizon ($1$, $3$ and $6$ months), the estimates are performed under two model specifications: the true model and the best fitting model based on BIC, denoted, respectively,  with the superscripts $TRUE$ and $BIC$.}
\label{tab:sim_res_coverage}
\scalebox{0.8}{
\begin{tabular}{@{}lrrrrrrr@{}}
\toprule
     & \multicolumn{3}{c}{C-ARIMA$^{BIC}$}  &  & \multicolumn{3}{c}{REG-ARIMA$^{BIC}$}  \\ \cmidrule{2-4} \cmidrule{6-8}
 $\tau_t(1;0)$    & 1 month   & 3 months  & 6 months &  & 1 month   & 3 months & 6 months \\ \midrule
$+1\%$   & 94.25 & 93.68 & 93.15 & & 95.20 & 94.78 & 95.62 \\ 
$+10\%$  & 94.25 & 93.68 & 93.15 & & 90.74 & 90.40 & 91.25 \\ 
$+25\%$  & 94.25 & 93.68 & 93.15 & & 71.89 & 68.01 & 69.44 \\ 
$+50\%$  & 94.25 & 93.68 & 93.15 & & 45.62 & 40.66 & 42.59 \\ 
$+100\%$ & 94.25 & 93.68 & 93.15 & & 24.83 & 23.91 & 27.02 \\ 
NS       & 94.21 & 93.66 & 93.16 & & 0.17 & 0.00 & 0.00 \\ 
\midrule
     & \multicolumn{3}{c}{C-ARIMA$^{TRUE}$} &  & \multicolumn{3}{c}{REG-ARIMA$^{TRUE}$} \\ \cmidrule{2-4} \cmidrule{6-8}
 $\tau_t(1;0)$    & 1 month   & 3 months  & 6 months &  & 1 month   & 3 months & 6 months \\ \midrule
$+1\%$   & 94.27 & 93.70 & 93.19 & & 95.12 & 94.95 & 95.71 \\ 
$+10\%$  & 94.27 & 93.70 & 93.19 & & 90.99 & 90.40 & 91.67 \\ 
$+25\%$  & 94.27 & 93.70 & 93.19 & & 71.89 & 67.93 & 69.70 \\ 
$+50\%$  & 94.27 & 93.70 & 93.19 & & 45.62 & 40.74 & 43.10 \\ 
$+100\%$ & 94.27 & 93.70 & 93.19 & & 24.92 & 23.99 & 27.02 \\ 
NS       & 94.23 & 93.69 & 93.20 & & 0.17 & 0.00 & 0.00 \\ 
\bottomrule
\end{tabular}

}
\end{table}

\clearpage
\section{Empirical application}
\label{sect:data_method}
In this section we describe the results of our empirical application; the goal is to estimate the impact of the permanent price reduction performed by an Italian supermarket chain.

\subsection{Data and methodology}

\begin{figure}[h!]
\caption{Daily time series of unit sold, price per unit and autocorrelation function for two selected items (i.e., store brand $6$ and the corresponding competitor brand). For the competitor, the relative price plot shows the ratio between its unit price and the price of the store brand.}
\label{fig:des_ex}
\begin{tabular}{m{2.3cm}m{9cm}}
\scriptsize \textbf{Store} & \includegraphics[scale=0.55]{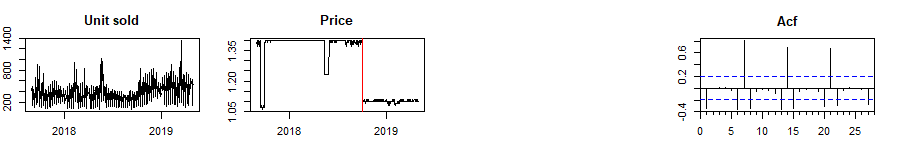} \\
\scriptsize \textbf{Competitor} & \includegraphics[scale=0.55]{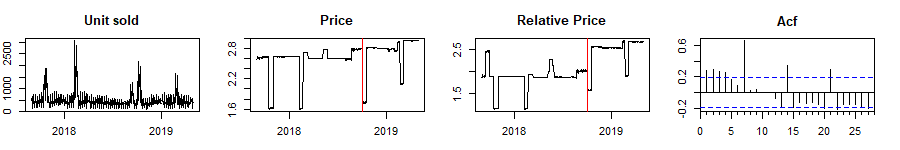} \\
\end{tabular}
\end{figure}

Data consists of daily sales counts of $11$ store brands and their corresponding competitor brand cookies in the period September 1, 2017, April 30, 2019.\footnote{We excluded the last competitor brand because $62 \%$ of observations were missing. Thus, we analyzed $11$ store and $10$ competitor brands.} The permanent price reduction on the store brand cookies was introduced by the supermarket chain on October 4, 2018.
As an example, Figure \ref{fig:des_ex} shows the time series of units sold, the evolution of price per unit and the autocorrelation function of one store brand and its direct competitor. The plots for the remaining store-brand and competitor-brand cookies are provided in Appendix \ref{appA}. The occasional price drops before the intervention date indicate temporary promotions run regularly by the supermarket chain. The products exhibit a clear weekly seasonal pattern, evidenced by the spikes in the autocorrelation functions. In the panel referred to the direct competitor brand, we can also observe the evolution of the relative price per unit (the ratio between the prices of the competitor brand and the corresponding store brand).  Unsurprisingly, despite the occasional drops due to temporary promotions, the price of the competitor brand relative to the corresponding store brand has increased after the intervention. 

To estimate the causal effect of the permanent price discount on the sales of store-brands cookies, we follow the approach outlined in Section \ref{sect:cARIMA}. In particular, under Assumption \ref{assumpt:no-interference}, we analyze each cookie separately, thereby fitting $11$ independent models. In order to improve model diagnostics, the dependent variable is the natural log of the daily sales count. This also means that we are postulating the existence of a multiplicative effect of the new price policy on the sales of cookies. Since in terms of the original variable the cumulative sum of daily effects is equivalent to their product, we focused our attention on estimating the temporal average causal effect, which can still be interpreted as an average multiplicative effect. Furthermore, we included covariates to improve prediction of the missing potential outcomes in the absence of intervention. In particular, to take care of the seasonality we included six dummy variables corresponding to the day of the week and one dummy denoting December Sundays.\footnote{In principle, we may also have a monthly seasonal pattern on top of the weekly cycle but the reduced length of the pre-intervention time series ($398$ observations) does not allow us to assess whether a double seasonality is present.} Indeed, the policy of the supermarket chain implies that all shops are closed on Sunday afternoon except during Christmas holidays. Thus, we may have two opposite ``Sunday effects'': a positive effect in December, when the shops are open all the day; a negative effect during the rest of the year, since all shops are closed in the afternoon. We also included a holiday dummy taking value $1$ before and after a national holiday and $0$ otherwise. This is to account for consumers' tendency to increase purchases before and after a closure day.\footnote{To be precise, on the day of a national holiday we have a missing value (so there is no holiday effect), whereas the dummy variable should capture the effect of additional purchases before and after the closure day(s).} Finally, we included a modified version of the unit price, that after the intervention day and during all the post-period is taken equal to the last price before the permanent discount. As explained in discussing Assumption \ref{assumpt:covariates}, this is the most likely price that the unit would have had in the absence of intervention. In addition, to estimate the average causal effect of the intervention on store brands, we are also interested in evaluating how this effect evolves with time. Thus, we repeated the analysis by making predictions at three different time horizons: $1$ month, $3$ months and $6$ months after the intervention.

The same methodology is applied to the competitor brands, with a slight modification on the set of covariates. Indeed, this time the unit price is not directly influenced by the intervention, which instead affects the relative price (as shown in Figure \ref{fig:des_ex}); so, to forecast competitor sales in the absence of intervention we directly used the actual price. 

Again, to illustrate the merits of the proposed approach, the results obtained from C-ARIMA are then compared to those of REG-ARIMA, as described by Equation (\ref{eqn:ARIMA_errors}). More specifically, we fitted independent linear regressions with ARIMA errors for each of the $11$ store brands and their competitors. 

\subsection{Results and discussion}
\label{subsect:results}

\begin{figure}[h!]
\centering
\caption{First row: observed sales (gray) and forecasted sales (blue) of store brand 4 at $1$ month (horizon 1), $3$ months (horizon 2) and $6$ months (horizon 3) from the intervention; the vertical bar indicates the intervention date. Second row: pattern of the point causal effect, computed as the difference between observed and forecasted sales, with its 95\% confidence interval.}
\label{fig:tp_arima_causal_ex}
\includegraphics[width = 16cm, height = 7cm]{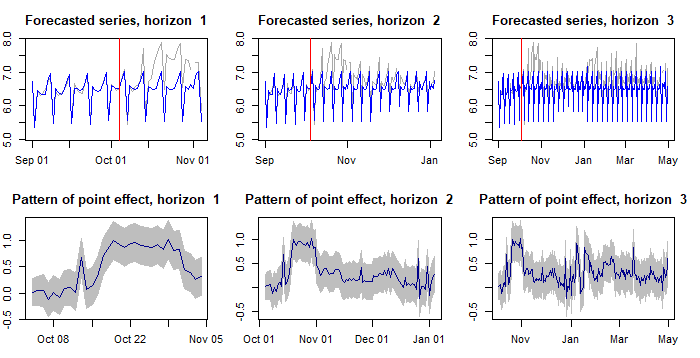} \\
\end{figure}

Table \ref{tab:tp_arima_causal} shows the results of the C-ARIMA and the REG-ARIMA approaches applied to the store brands.\footnote{For the C-ARIMA approach, Table \ref{tab:boot_store} shows that the results with the empirical critical values computed from the bootstrapped errors are in line with those reported in Table \ref{tab:tp_arima_causal}.} Figure \ref{fig:tp_arima_causal_ex} illustrates the causal effect, the observed time series and the forecasted series in the absence of intervention for one selected item.\footnote{The same plots for the remaining store-brand and competitor-brand cookies are provided, respectively, in Appendix \ref{appA_store} and Appendix \ref{appA_competitor}.} At the $1$-month time horizon, the causal effect is significantly positive for $8$ out of $11$ items; three months after the intervention, the causal effect is significantly positive for $10$ items; after six months, the effect is significant and positive for all items. Conversely, REG-ARIMA fails to detect some of the effects: compared to the C-ARIMA results, the effect on items $4$ and $11$ at the first time horizon, on item $11$ at the second horizon and on items $5$ and $11$ at the third horizon are not significant.  

\begin{figure}[h!]
\centering
\caption{First row: observed sales (grey) and forecasted sales (blue) of competitor brand 10 at $1$ month (horizon 1), $3$ months (horizon 2) and $6$ months (horizon 3) from the intervention; the vertical bar indicates the intervention date. Second row: pattern of the point causal effect, computed as the difference between observed and forecasted sales, with its 95\% confidence interval.}
\label{fig:sp_arima_causal_ex}
\includegraphics[width = 16cm, height = 6.5cm]{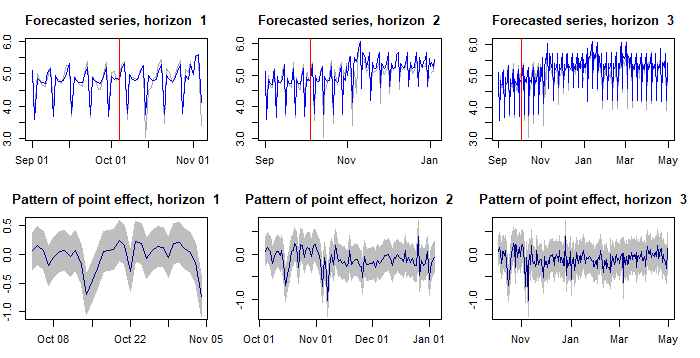}  

\end{figure}

Table \ref{tab:sp_arima_causal} reports the results for the competitor brands and Figure \ref{fig:sp_arima_causal_ex} plots the causal effect, the observed series and the forecasted series for one selected item.\footnote{For the C-ARIMA approach, Table \ref{tab:boot_comp} shows that the results with the empirical critical values computed from the bootstrapped errors are in line with those reported in Table \ref{tab:sp_arima_causal}.} Again, the causal effect seems to strengthen as we proceed far away from the intervention. At $1$-month horizon no significant effect is observed; three months after the intervention, we find a significant and negative effect on item $10$; at $6$-month horizon we find significant negative effects on items $8$ and $10$ and a significant positive effect on item $5$. A negative effect suggests that following the permanent price discount, consumers have changed their behavior by privileging the cheaper store brand. Instead, a positive effect might indicate that the price policy has determined an increase in the customer base, i.e. new clients have entered the shop and eventually bought the items at full price. Again, REG-ARIMA model leads to partially different results: at $6$-month horizon, a positive effect is found on item $6$ and no effect is detected on item $8$.

Summarizing, the intervention seems to have produced a significant and positive effect on the sales of store brand cookies. Conversely, we do not find considerable evidence of a detrimental effect on competitor cookies (the only exceptions being items $8$ and $10$). This indicates that, even though each store-competitor pair is formed by perfect substitutes, price might not be the only factor driving sales. For example, unobserved factors such as individual preferences or brand faithfulness may have a role as well. 

\begin{table}[h!]
\centering
\caption{Causal effect estimates of the permanent price rebate on sales of store-brand cookies after one month, three months and six months from the intervention. In this table, $ \hat{\bar{\tau}}^Y_t $ is the estimated temporal average effect on the original variable ($ \hat{\bar{\tau}}^Y_t = 0 $ implies no effect), while $ \hat{\beta}_0$ is the coefficient estimate of the intervention dummy according to REG-ARIMA ($ \hat{\beta}_0 = 0 $ implies absence of association). Standard errors within parentheses.}
\label{tab:tp_arima_causal}
\scalebox{0.8}{
\begin{tabular}{lr D{.}{.}{-2} D{.}{.}{-2} r D{.}{.}{-2} D{.}{.}{-2} r D{.}{.}{-2} D{.}{.}{-2}} 
\\[-1.8ex]\hline 
\hline \\[-1.8ex] 
& \multicolumn{9}{c}{\textit{Time horizon:}} \\ 
\cline{2-10}
\\[-1.8ex]   &  & \multicolumn{2}{c}{$1$ month} &  & \multicolumn{2}{c}{$3$ months} &  & \multicolumn{2}{c}{$6$ months} \\                      
Item                &  & \hat{\bar{\tau}}_t   & \hat{\beta}_0    &  & \hat{\bar{\tau}}_t          &\hat{\beta}_0     &  & \hat{\bar{\tau}}_t          & \hat{\beta}_0     \\ \hline \\[-1.8ex] 
\multirow{2}{*}{1}  &  & 0.14      & 0.14      & & 0.15^{\bs{.}}    &  0.12^{\bs{.}} & & 0.18^{***} & 0.16^{**}  \\
                    &  & (0.12)    & (0.09)    & & (0.08)      &  (0.07)   & &  (0.06)    & (0.06) \\
\multirow{2}{*}{2}  &  & 0.14      &  0.10      & & 0.13^{\bs{.}}    &  0.12^{\bs{.}} & & 0.14^{**}  & 0.13^{**}   \\
                    &  & (0.12)    & (0.14)    & & (0.08)      &  (0.07)   & &  (0.05)    & (0.05) \\
\multirow{2}{*}{3}  &  & 0.19^{\bs{.}}  & 0.15^{\bs{.}}  & & 0.21^{**}   &  0.15^{*} & & 0.25^{***} & 0.24^{***}  \\
                    &  & (0.11)    & (0.08)    & & (0.07)      &  (0.07)   & &  (0.05)    & (0.04)         \\
\multirow{2}{*}{4}  &  & 0.49^{***}&  0.00     & & 0.30^{***}   &  0.19^{*} & & 0.32^{***} & 0.28^{***}  \\
                    &  & (0.09)    & (0.13)    & & (0.06)      &  (0.08)   & &  (0.04)    & (0.05)        \\
\multirow{2}{*}{5}  &  &  -0.02    &  -0.06    & &   0.07      &  -0.07    & & 0.11^{\bs{.}}   &   -0.06         \\
                    &  & (0.12)    & (0.12)    & & (0.08)      &  (0.11)   & &  (0.06)    &  (0.10)          \\
\multirow{2}{*}{6}  &  & 0.24^{*}  & 0.26^{\bs{.}}  & &0.34^{***}   &  0.24^{*} & & 0.37^{***} &  0.23^{*}   \\
                    &  & (0.12)    & (0.14)    & & (0.08)      & (0.12)    & & (0.06)     & (0.11)         \\
\multirow{2}{*}{7}  &  & 0.55^{***}& 0.75^{***}& &  0.34^{***} &  0.70^{***}& &  0.30^{***} & 0.77^{***}  \\
                    &  & (0.10)     &(0.11)     & & (0.07)      & (0.10)     & & (0.05)     & (0.11)         \\
\multirow{2}{*}{8}  &  & 0.26^{***}&  0.29^{**}& & 0.25^{***}  & 0.29^{**} & & 0.14^{**}  & 0.28^{**}  \\
                    &  & (0.08)    & (0.09)    & & (0.07)      & (0.10)     & & (0.05)     & (0.09)         \\
\multirow{2}{*}{9}  &  & 0.47^{***}&  0.70^{***}& &  0.20^{***}  & 0.29^{***}& & 0.21^{***} & 0.26^{***}  \\
                    &  & (0.06)    &  (0.10)    & & (0.04)      & (0.09)    & & (0.03)     & (0.06)         \\
\multirow{2}{*}{10} &  & 0.66^{***}& 0.85^{***}& & 0.57^{***}  & 0.82^{***}& & 0.33^{***} & 0.85^{***}  \\
                    &  & (0.11)    & (0.14)    & & (0.08)      & (0.15)    & & (0.06)     & (0.15)         \\
\multirow{2}{*}{11} &  & 0.12^{*}  &     0.02  & & 0.16^{**}   &  0.04     & & 0.14^{***} &  0.08          \\
                    &  & (0.06)    & (0.11)    & & (0.05)      & (0.11)    & & (0.04)     & (0.13)        \\ 
\hline
\hline \\[-1.8ex]
\textit{Note:}      & \multicolumn{9}{r}{$^{\boldsymbol{\cdot}}$p$<$0.1; $^{*}$p$<$0.05; $^{**}$p$<$0.01; $^{***}$p$<$0.001} \\                  
\end{tabular}
}
\end{table}

\clearpage

\begin{table}[h!]
\centering
\caption{Causal effect estimates of the permanent price rebate on sales of competitor-brand cookies after one month, three months and six months from the intervention. In this table, $ \hat{\bar{\tau}}^Y_t$ is the estimated temporal average effect on the original variable ($ \hat{\bar{\tau}}^Y_t = 0 $ implies no effect), while $ \hat{\beta}_0$ is the coefficient estimate of the intervention dummy according to REG-ARIMA ($ \hat{\beta}_0 = 0 $ implies absence of association). Standard errors within parentheses.}
\label{tab:sp_arima_causal}
\scalebox{0.8}{
\begin{tabular}{lr D{.}{.}{-2} D{.}{.}{-2} r D{.}{.}{-2} D{.}{.}{-2} r D{.}{.}{-2} D{.}{.}{-2}} 
\\[-1.8ex]\hline 
\hline \\[-1.8ex] 
& \multicolumn{9}{c}{\textit{Time horizon:}} \\ 
\cline{2-10}
\\[-1.8ex]   &  & \multicolumn{2}{c}{$1$ month} &  & \multicolumn{2}{c}{$3$ months} &  & \multicolumn{2}{c}{$6$ months} \\                      
Item                &  & \hat{\bar{\tau}}_t   & \hat{\beta}_0    &  & \hat{\bar{\tau}}_t          &\hat{\beta}_0     &  & \hat{\bar{\tau}}_t          & \hat{\beta}_0     \\ \hline \\[-1.8ex] 
\multirow{2}{*}{1}  &  &  -0.04 & 0.02   & & 0.02       & -0.16     & & 0.04        & -0.12      \\
                    &  & (0.56) & (0.19) & & (0.46)     & (0.25)    & & (0.34)      & (0.22)     \\
\multirow{2}{*}{2}  &  &  -0.13 & -0.18  & & -0.07      & -0.13     & & -0.15       & -0.13      \\
                    &  & (0.50) & (0.22) & & (0.47)     & (0.20)    & & (0.36)      & (0.19)     \\
\multirow{2}{*}{3}  &  & 0.04   & -0.06  & & 0.09       & -0.03     & & 0.17        & 0.03       \\        
                    &  & (0.38) & (0.22) & & (0.23)     & (0.20)    & & (0.11)      & (0.17)     \\
\multirow{2}{*}{4}  &  &  0.00  & 0.08   & & -0.13      & 0.02      & & -0.04       & 0.01       \\
                    &  & (0.29) & (0.21) & & (0.21)     & (0.22)    & & (0.14)      & (0.13)     \\
\multirow{2}{*}{5}  &  & -0.03  & -0.01  & & 0.05       & 0.06      & & 0.12^{**}   & 0.12^{*}   \\
                    &  & (0.10) & (0.10) & & (0.06)     & (0.06)    & & (0.04)      & (0.05)     \\
\multirow{2}{*}{6}  &  & -0.05  & -0.01  & & 0.03       & 0.06      & & 0.09        & 0.10^{*}   \\
                    &  & (0.12) & (0.10) & & (0.09)     & (0.06)    & & (0.07)      & (0.05)     \\
\multirow{2}{*}{7}  &  & 0.04   & -0.11  & & 0.11       & -0.05     & & 0.40^{\bs{.}}    & 0.02  \\
                    &  & (0.54) & (0.29) & & (0.33)     & (0.26)    & & (0.23)      & (0.23)     \\
\multirow{2}{*}{8}  &  &  -0.09 & -0.02  & & -0.06      & -0.10     & & -0.08^{*}   & -0.12      \\
                    &  & (0.07) & (0.07) & & (0.05)     & (0.10)    & & (0.04)      & (0.10)     \\
\multirow{2}{*}{9}  &  &  -0.09 & -0.08  & & -0.11      & -0.11     & & -0.10       & -0.09      \\
                    &  & (0.13) & (0.13) & & (0.09)     & (0.08)    & & (0.06)      & (0.06)     \\
\multirow{2}{*}{10} &  & -0.03  & -0.02  & & -0.12^{**} & -0.09^{*} & & -0.11^{***} & -0.08^{*}  \\
                    &  & (0.06) & (0.05) & & (0.04)     & (0.04)    & & (0.03)      & (0.04)     \\
\hline
\hline \\[-1.8ex]
\textit{Note:}      & \multicolumn{9}{r}{$^{\boldsymbol{\cdot}}$p$<$0.1; $^{*}$p$<$0.05; $^{**}$p$<$0.01; $^{***}$p$<$0.001} \\                  
\end{tabular}
}
\end{table}

\section{Conclusions}
\label{sect:conclusions}
We propose a novel approach, C-ARIMA, to estimate the effect of interventions in a time series setting under the Rubin Causal Model. After a detailed illustration of the assumptions underneath our causal framework, we defined three causal estimands of interest, i.e., the point, cumulative and average causal effects. Then, we introduced a methodology to perform inference. 

To measure the performance of C-ARIMA in uncovering causal effects, we presented a simulation study showing that this approach performs well in comparison with a standard intervention analysis approach (REG-ARIMA) where the true effect is in the form of a level shift; it also outperforms the latter in case of irregular, time-varying effects. 

Finally, we applied the proposed methodology to estimate the causal effect of the new price policy introduced by a big supermarket chain in Italy, which addressed a selected subset of store-brands products by permanently lowering their price. The empirical analysis was carried out on the goods belonging to the  ``cookies'' category: the results show that the permanent price reduction was effective in increasing store-brand cookies' sales. Little evidence of a detrimental effect on the corresponding competitor-brand cookies is found.   

\bibliography{C:/Users/fiamm/Dropbox/references}

\begin{thebibliography}{}

\bibitem[Abadie, 2005]{Abadie:2005}
Abadie, A. (2005).
\newblock Semiparametric difference-in-differences estimators.
\newblock {\em The Review of Economic Studies}, 72(1):1--19.

\bibitem[Abadie et~al., 2010]{Abadie:Diamond:Hainmueller:2010}
Abadie, A., Diamond, A., and Hainmueller, J. (2010).
\newblock Synthetic control methods for comparative case studies: Estimating
  the effect of california’s tobacco control program.
\newblock {\em Journal of the American Statistical Association},
  105(490):493--505.

\bibitem[Abadie et~al., 2015]{Abadie:Diamond:Hainmueller:2015}
Abadie, A., Diamond, A., and Hainmueller, J. (2015).
\newblock Comparative politics and the synthetic control method.
\newblock {\em American Journal of Political Science}, 59(2):495--510.

\bibitem[Abadie and Gardeazabal, 2003]{Abadie:Gardeazabal:2003}
Abadie, A. and Gardeazabal, J. (2003).
\newblock The economic costs of conflict: A case study of the basque country.
\newblock {\em American economic review}, 93(1):113--132.

\bibitem[Anger et~al., 2011]{Anger:Kvasnicka:Siedler:2011}
Anger, S., Kvasnicka, M., and Siedler, T. (2011).
\newblock One last puff? public smoking bans and smoking behavior.
\newblock {\em Journal of health economics}, 30(3):591--601.

\bibitem[Angrist and Pischke, 2008]{Angrist:Pischke:2008}
Angrist, J.~D. and Pischke, J.-S. (2008).
\newblock {\em Mostly harmless econometrics: An empiricist's companion}.
\newblock Princeton university press.

\bibitem[Arkhangelsky et~al.,
  2019]{Arkhangelsky:Athey:Hirshberg:Imbens:Wager:2019}
Arkhangelsky, D., Athey, S., Hirshberg, D.~A., Imbens, G.~W., and Wager, S.
  (2019).
\newblock Synthetic difference in differences.
\newblock Technical report, National Bureau of Economic Research.

\bibitem[Athey and Imbens, 2021]{Athey:Imbens:2021}
Athey, S. and Imbens, G.~W. (2021).
\newblock Design-based analysis in difference-in-differences settings with
  staggered adoption.
\newblock {\em Journal of Econometrics}.

\bibitem[Balke and Fomby, 1994]{Balke:Fomby:1994}
Balke, N.~S. and Fomby, T.~B. (1994).
\newblock Large shocks, small shocks, and economic fluctuations: Outliers in
  macroeconomic time series.
\newblock {\em Journal of Applied Econometrics}, 9(2):181--200.

\bibitem[Ben-Michael et~al., 2018]{Benmichael:Feller:Rothstein:2018}
Ben-Michael, E., Feller, A., and Rothstein, J. (2018).
\newblock The augmented synthetic control method.
\newblock Preprint. Available at \url{arXiv:1811.04170}.

\bibitem[Bhattacharyya and Layton, 1979]{Bhattacharyya:Layton:1979}
Bhattacharyya, M. and Layton, A.~P. (1979).
\newblock Effectiveness of seat belt legislation on the queensland road
  toll—an {A}ustralian case study in intervention analysis.
\newblock {\em Journal of the American Statistical Association},
  74(367):596--603.

\bibitem[Billmeier and Nannicini, 2013]{Billmeier:Nannicini:2013}
Billmeier, A. and Nannicini, T. (2013).
\newblock Assessing economic liberalization episodes: A synthetic control
  approach.
\newblock {\em Review of Economics and Statistics}, 95(3):983--1001.

\bibitem[Bojinov et~al., 2020]{Bojinov:Rambachan:Shephard:2020}
Bojinov, I., Rambachan, A., and Shephard, N. (2020).
\newblock Panel experiments and dynamic causal effects: A finite population
  perspective.
\newblock Preprint. Available at \url{arXiv:2003.09915}.

\bibitem[Bojinov and Shephard, 2019]{Bojinov:Shephard:2019}
Bojinov, I. and Shephard, N. (2019).
\newblock Time series experiments and causal estimands: exact randomization
  tests and trading.
\newblock {\em Journal of the American Statistical Association},
  114(528):1665--1682.

\bibitem[Box and Tiao, 1975]{Box:Tiao:1975}
Box, G.~E. and Tiao, G.~C. (1975).
\newblock Intervention analysis with applications to economic and environmental
  problems.
\newblock {\em Journal of the American Statistical Association},
  70(349):70--79.

\bibitem[Box and Tiao, 1976]{Box:Tiao:1976}
Box, G.~E. and Tiao, G.~C. (1976).
\newblock Comparison of forecast and actuality.
\newblock {\em Journal of the Royal Statistical Society: Series C (Applied
  Statistics)}, 25(3):195--200.

\bibitem[Brodersen et~al., 2015]{Brodersen:Gallusser:Koehler:Remy:Scott:2015}
Brodersen, K.~H., Gallusser, F., Koehler, J., Remy, N., and Scott, S.~L.
  (2015).
\newblock Inferring causal impact using {B}ayesian structural time-series
  models.
\newblock {\em The Annals of Applied Statistics}, 9(1):247--274.

\bibitem[Callaway and Sant’Anna, 2020]{Callaway:Santanna:2020}
Callaway, B. and Sant’Anna, P.~H. (2020).
\newblock Difference-in-differences with multiple time periods.
\newblock {\em Journal of Econometrics}.

\bibitem[Card and Krueger, 1993]{Card:Krueger:1993}
Card, D. and Krueger, A.~B. (1993).
\newblock Minimum wages and employment: A case study of the fast food industry
  in new jersey and pennsylvania.
\newblock Technical report, National Bureau of Economic Research.

\bibitem[Cauley and Im, 1988]{Cauley:Iksoon:1988}
Cauley, J. and Im, E.~I. (1988).
\newblock Intervention policy analysis of skyjackings and other terrorist
  incidents.
\newblock {\em The American Economic Review}, 78(2):27--31.

\bibitem[Chen and Liu, 1993]{Chen:Liu:1993}
Chen, C. and Liu, L.-M. (1993).
\newblock Joint estimation of model parameters and outlier effects in time
  series.
\newblock {\em Journal of the American Statistical Association}, 88.

\bibitem[Ding and Li, 2019]{Ding:Li:2019}
Ding, P. and Li, F. (2019).
\newblock A bracketing relationship between difference-in-differences and
  lagged-dependent-variable adjustment.
\newblock {\em Political Analysis}, 27(4):605--615.

\bibitem[Dube and Zipperer, 2015]{Dube:Zipperer:2015}
Dube, A. and Zipperer, B. (2015).
\newblock Pooling multiple case studies using synthetic controls: An
  application to minimum wage policies.
\newblock {\em IZA Discussion Paper 8944}.

\bibitem[Enders and Sandler, 1993]{Enders:Sandler:1993}
Enders, W. and Sandler, T. (1993).
\newblock The effectiveness of antiterrorism policies: A
  {V}ector-{A}uto{R}egression-intervention analysis.
\newblock {\em The American Political Science Review}, 87(4):829--844.

\bibitem[Forastiere et~al., 2020]{Forastiere:Airoldi:Mealli:2020}
Forastiere, L., Airoldi, E.~M., and Mealli, F. (2020).
\newblock Identification and estimation of treatment and interference effects
  in observational studies on networks.
\newblock {\em Journal of the American Statistical Association}.

\bibitem[Garvey and Hanka, 1999]{Garvey:Hanka:1999}
Garvey, G.~T. and Hanka, G. (1999).
\newblock Capital structure and corporate control: The effect of antitakeover
  statutes on firm leverage.
\newblock {\em The Journal of Finance}, 54(2):519--546.

\bibitem[Gobillon and Magnac, 2016]{Gobillon:Magnac:2016}
Gobillon, L. and Magnac, T. (2016).
\newblock Regional policy evaluation: Interactive fixed effects and synthetic
  controls.
\newblock {\em Review of Economics and Statistics}, 98(3):535--551.

\bibitem[Harvey, 1989]{Harvey:1989}
Harvey, A.~C. (1989).
\newblock {\em Forecasting, structural time series models and the Kalman
  filter}.
\newblock Cambridge university press.

\bibitem[Holland, 1986]{Holland:1986}
Holland, P.~W. (1986).
\newblock Statistics and causal inference.
\newblock {\em Journal of the American statistical Association},
  81(396):945--960.

\bibitem[Imbens and Rubin, 2015]{Imbens:Rubin:2015}
Imbens, G.~W. and Rubin, D.~B. (2015).
\newblock {\em Causal inference in Statistics, Social, and Biomedical
  Sciences}.
\newblock Cambridge University Press.

\bibitem[Kreif et~al.,
  2016]{Kreif:Grieve:Hangartner:Turner:Nikolova:Sutton:2016}
Kreif, N., Grieve, R., Hangartner, D., Turner, A.~J., Nikolova, S., and Sutton,
  M. (2016).
\newblock Examination of the synthetic control method for evaluating health
  policies with multiple treated units.
\newblock {\em Health economics}, 25(12):1514--1528.

\bibitem[Larcker et~al., 1980]{Larcker:Gordon:1980}
Larcker, D.~F., Gordon, L.~A., and Pinches, G.~E. (1980).
\newblock Testing for market efficiency: a comparison of the cumulative average
  residual methodology and intervention analysis.
\newblock {\em Journal of Financial and Quantitative Analysis}, 15(2):267--287.

\bibitem[Li, 2019]{Li:2019}
Li, K.~T. (2019).
\newblock Statistical inference for average treatment effects estimated by
  synthetic control methods.
\newblock {\em Journal of the American Statistical Association}.

\bibitem[Menchetti and Bojinov, 2020]{Menchetti:Bojinov:2020}
Menchetti, F. and Bojinov, I. (2020).
\newblock Estimating causal effects in the presence of partial interference
  using multivariate bayesian structural time series models.
\newblock {\em Preprint. Available at \url{arXiv:2006.12269}}.

\bibitem[Meyer et~al., 1995]{Meyer:Viscusi:Durbin:1995}
Meyer, B.~D., Viscusi, W.~K., and Durbin, D.~L. (1995).
\newblock Workers' compensation and injury duration: evidence from a natural
  experiment.
\newblock {\em The American economic review}, 85(3):322--340.

\bibitem[Murry et~al., 1993]{Murry:Stam:Lastovicka:1993}
Murry, J.~P., Stam, A., and Lastovicka, J.~L. (1993).
\newblock Evaluating an anti-drinking and driving advertising campaign with a
  sample survey and time series intervention analysis.
\newblock {\em Journal of the American Statistical Association},
  88(421):50--56.

\bibitem[Noirjean et~al., 2020]{Noirjean:Mariani:Mattei:Mealli:2020}
Noirjean, S., Mariani, M., Mattei, A., and Mealli, F. (2020).
\newblock Exploiting network information to disentangle spillover effects in a
  field experiment on teens' museum attendance.
\newblock {\em arXiv preprint arXiv:2011.11023}.

\bibitem[O’Neill et~al., 2016]{ONeill:Kreif:Grieve:Sutton:Sekhon:2016}
O’Neill, S., Kreif, N., Grieve, R., Sutton, M., and Sekhon, J.~S. (2016).
\newblock Estimating causal effects: considering three alternatives to
  difference-in-differences estimation.
\newblock {\em Health Services and Outcomes Research Methodology},
  16(1-2):1--21.

\bibitem[Papadogeorgou et~al.,
  2018]{Choirat:Dominici:Mealli:Papadogeorgou:Wasfy:Zigler:2018}
Papadogeorgou, G., Mealli, F., Zigler, C.~M., Dominici, F., Wasfy, J.~H., and
  Choirat, C. (2018).
\newblock Causal impact of the hospital readmissions reduction program on
  hospital readmissions and mortality.
\newblock Preprint. Available at \url{arXiv:1809.09590}.

\bibitem[Rambachan and Roth, 2019]{Rambachan:Roth:2019}
Rambachan, A. and Roth, J. (2019).
\newblock An honest approach to parallel trends.
\newblock {\em Unpublished manuscript, Harvard University.}

\bibitem[Rambachan and Shephard, 2019]{Rambachan:Shephard:2019}
Rambachan, A. and Shephard, N. (2019).
\newblock A nonparametric dynamic causal model for macroeconometrics.
\newblock {\em Preprint. Available at \url{arXiv preprint arXiv:1903.01637}}.

\bibitem[Robins, 1986]{Robins:1986}
Robins, J.~M. (1986).
\newblock A new approach to causal inference in mortality studies with a
  sustained exposure period—application to control of the healthy worker
  survivor effect.
\newblock {\em Mathematical modelling}, 7(9-12):1393--1512.

\bibitem[Robins et~al., 1999]{Robins:Greenland:Hu:1999}
Robins, J.~M., Greenland, S., and Hu, F.-C. (1999).
\newblock Estimation of the causal effect of a time-varying exposure on the
  marginal mean of a repeated binary outcome.
\newblock {\em Journal of the American Statistical Association},
  94(447):687--700.

\bibitem[Roth, 2018]{Roth:2018}
Roth, J. (2018).
\newblock Should we adjust for the test for pre-trends in
  difference-in-difference designs?
\newblock {\em arXiv preprint arXiv:1804.01208}.

\bibitem[Rubin, 1974]{Rubin:1974}
Rubin, D.~B. (1974).
\newblock Estimating causal effects of treatments in randomized and
  nonrandomized studies.
\newblock {\em Journal of Educational Psychology}, 66(5):688--701.

\bibitem[Rubin, 1975]{Rubin:1975}
Rubin, D.~B. (1975).
\newblock Bayesian inference for causality: The importance of randomization.
\newblock In {\em The Proceedings of the Social Statistics Section of the
  American Statistical Association}, pages 233--239.

\bibitem[Rubin, 1978]{Rubin:1978}
Rubin, D.~B. (1978).
\newblock Bayesian inference for causal effects: The role of randomization.
\newblock {\em The Annals of Statistics}, 6(1):34--58.

\bibitem[Ryan et~al., 2015]{Ryan:Burgess:Dimick:2015}
Ryan, A.~M., Burgess~Jr, J.~F., and Dimick, J.~B. (2015).
\newblock Why we should not be indifferent to specification choices for
  difference-in-differences.
\newblock {\em Health services research}, 50(4):1211--1235.

\bibitem[Sun and Abraham, 2020]{Sun:Abraham:2020}
Sun, L. and Abraham, S. (2020).
\newblock Estimating dynamic treatment effects in event studies with
  heterogeneous treatment effects.
\newblock {\em Journal of Econometrics}.

\bibitem[VanderWeele, 2010]{VanderWeele:2010}
VanderWeele, T.~J. (2010).
\newblock Direct and indirect effects for neighborhood-based clustered and
  longitudinal data.
\newblock {\em Sociological methods \& research}, 38(4):515--544.

\bibitem[Viviano and Bradic, 2019]{Viviano:Bradic:2019}
Viviano, D. and Bradic, J. (2019).
\newblock Synthetic learner: model-free inference on treatments over time.
\newblock Preprint. Available at \url{arXiv:1904.01490}.

\bibitem[West and Harrison, 2006]{West:Harrison:2006}
West, M. and Harrison, J. (2006).
\newblock {\em Bayesian forecasting and dynamic models}.
\newblock Springer Science \& Business Media.

\end{thebibliography}

\newpage
\begin{appendices}

Appendix \ref{appA_store} includes additional tables and plots referred to the empirical analysis on the store brands. Appendix \ref{appA_competitor} includes additional tables and plots referred to the empirical analysis on the competitor brands. Appendix \ref{appB_proof} presents the proof of Theorem \ref{theo} and Appendix \ref{appB_untransformed} presents the proof of Theorem \ref{theo_2}.

\section{}
\label{appA}
\numberwithin{figure}{section}
\numberwithin{table}{section}
\setcounter{figure}{0}
\setcounter{table}{0}

\subsection{Store brands}
\label{appA_store}

\begin{table}[h!]
    \centering
    \caption{Causal effect estimates of the permanent price rebate on sales of store-brand cookies after one month, three months and six months from the intervention. In this table, $ \hat{\bar{\tau}}^Y_t$ is the estimated temporal average effect on the original variable ($ \hat{\bar{\tau}}^Y_t = 0 $ implies no effect) and the empirical critical values are computed based on Equation (\ref{eqn:avetauhat-H0_2}) by bootstrapping the errors from the model residuals.}
    \label{tab:boot_store}
    \scalebox{0.8}{
    \begin{tabular}{l D{.}{.}{-2} r D{.}{.}{-2} r D{.}{.}{-2}} 
\\[-1.8ex]\hline 
\hline \\[-1.8ex] 
      & \multicolumn{5}{c}{\textit{Time horizon:}} \\ 
\cline{2-6}
\\[-1.8ex]     & $1 \text{ month}$  &  & $3 \text{ months}$  &  & $6 \text{ months}$  \\                      
Item  &  \hat{\bar{\tau}}_t   &  & \hat{\bar{\tau}}_t  &  & \hat{\bar{\tau}}_t  \\ \hline \\[-1.8ex] 
 1    &  0.14       &  &   0.15^{.}   &  & 0.18^{***} \\
 2    &  0.14       &  &   0.13^{.}   &  & 0.14^{*}   \\
 3    &  0.19^{.}   &  &   0.21^{**}  &  & 0.25^{***} \\
 4    &  0.49^{***} &  &   0.30^{***} &  & 0.32^{***} \\
 5    &  -0.02      &  &   0.07       &  & 0.11^{.}    \\
 6    &  0.24^{*}   &  &   0.34^{***} &  & 0.37^{***}  \\
 7    &  0.55^{***} &  &   0.34^{***} &  & 0.30^{***}  \\
 8    &  0.26^{**}  &  &   0.25^{***} &  & 0.14^{**}   \\
 9    &  0.47^{***} &  &   0.20^{***} &  & 0.21^{***}  \\
10    &  0.66^{***} &  &   0.57^{***} &  & 0.33^{***}  \\
11    &  0.12^{*}   &  &   0.16^{**}  &  & 0.14^{***}  \\
\hline
\hline \\[-1.8ex]
\textit{Note:}      & \multicolumn{5}{r}{$^{\boldsymbol{\cdot}}$p$<$0.1; $^{*}$p$<$0.05; $^{**}$p$<$0.01; $^{***}$p$<$0.001} \\                  
\end{tabular}
    }
\end{table}

\begin{sidewaystable}[h!]
\caption{Estimated coefficients (standard errors within parentheses) of the C-ARIMA models fitted to the $11$ store brands in the pre-intervention period. The dependent variable is the daily sales counts of each product in log scale.}
\scalebox{0.55}{
\begin{tabular}{@{\extracolsep{5pt}}lD{.}{.}{-3} D{.}{.}{-3} D{.}{.}{-3} D{.}{.}{-3} D{.}{.}{-3} D{.}{.}{-3} D{.}{.}{-3} D{.}{.}{-3} D{.}{.}{-3} D{.}{.}{-3} D{.}{.}{-3} } 
\\[-1.8ex]\hline 
\hline \\[-1.8ex] 
 & \multicolumn{11}{c}{\textit{Dependent variable:}} \\ 
\cline{2-12} 
\\[-1.8ex] & \multicolumn{1}{c}{Item 1} & \multicolumn{1}{c}{Item 2} & \multicolumn{1}{c}{Item 3} & \multicolumn{1}{c}{Item 4} & \multicolumn{1}{c}{Item 5} & \multicolumn{1}{c}{Item 6} & \multicolumn{1}{c}{Item 7} & \multicolumn{1}{c}{Item 8} & \multicolumn{1}{c}{Item 9} & \multicolumn{1}{c}{Item 10} & \multicolumn{1}{c}{Item 11} \\ 
\hline \\[-1.8ex] 
 $\phi_1$ & 0.863^{***} & 0.865^{***} & 0.833^{***} & 0.496^{***} & 0.867^{***} & 0.890^{***} & 0.828^{***} & 0.949^{***} & 0.839^{***} & 0.364^{***} & 0.951^{***} \\ 
  & (0.035) & (0.037) & (0.047) & (0.053) & (0.037) & (0.032) & (0.042) & (0.034) & (0.050) & (0.053) & (0.039) \\ 
  $\phi_2$ &  &  &  & 0.221^{***} &  &  &  &  &  & 0.177^{**} &  \\ 
  &  &  &  & (0.052) &  &  &  &  &  & (0.055) &  \\ 
  $\phi_3$ &  &  &  &  &  &  &  &  &  & 0.216^{***} &  \\ 
  &  &  &  &  &  &  &  &  &  & (0.054) &  \\
  $\theta_1$ & -0.323^{***} & -0.340^{***} & -0.348^{***} &  & -0.251^{**} & -0.458^{***} & -0.191^{*} & -0.509^{***} & -0.476^{***} &  & -0.841^{***} \\ 
  & (0.065) & (0.074) & (0.086) &  & (0.082) & (0.063) & (0.080) & (0.076) & (0.085) &  & (0.072) \\ 
  $\theta_2$ &  &  &  &  &  &  &  & -0.212^{**} &  &  &  \\ 
  &  &  &  &  &  &  &  & (0.074) &  &  &  \\    
  $\Phi_1$ & 0.120^{*} & 0.113^{*} & 0.203^{***} & 0.260^{***} & 0.235^{***} &  & 0.234^{***} & 0.240^{***} & 0.124^{*} & 0.774^{***} & 0.688^{***} \\ 
  & (0.056) & (0.057) & (0.055) & (0.053) & (0.056) &  & (0.057) & (0.059) & (0.059) & (0.204) & (0.146) \\ 
  $\Theta_1$ &  &  &  &  &  & 0.181^{**} &  &  &  & -0.682^{**} & -0.503^{**} \\ 
  &  &  &  &  &  & (0.058) &  &  &  & (0.231) & (0.172) \\ 
  c & 7.131^{***} & 6.767^{***} & 6.995^{***} & 7.912^{***} & 7.113^{***} & 6.535^{***} & 7.740^{***} & 7.800^{***} & 6.876^{***} & 6.312^{***} & 6.711^{***} \\ 
  & (0.167) & (0.167) & (0.159) & (0.152) & (0.072) & (0.107) & (0.099) & (0.063) & (0.025) & (0.110) & (0.044) \\ 
  price & -1.721^{***} & -1.441^{***} & -1.656^{***} & -2.218^{***} & -1.433^{***} & -1.841^{***} & -1.423^{***} & -2.316^{***} & -1.727^{***} & -2.118^{***} & -1.175^{***} \\ 
  & (0.356) & (0.355) & (0.339) & (0.288) & (0.287) & (0.304) & (0.281) & (0.116) & (0.117) & (0.176) & (0.177) \\ 
  hol & 0.176^{***} & 0.165^{***} & 0.168^{***} & 0.160^{***} & 0.174^{***} & 0.162^{***} & 0.162^{***} & 0.164^{***} & 0.163^{***} & 0.076^{*} & 0.208^{***} \\ 
  & (0.033) & (0.033) & (0.034) & (0.029) & (0.027) & (0.033) & (0.027) & (0.023) & (0.026) & (0.035) & (0.028) \\ 
  Dec.Sun & 0.323^{***} & 0.369^{***} & 0.306^{***} & 0.292^{***} & 0.379^{***} & 0.411^{***} & 0.367^{***} & 0.327^{***} & 0.390^{***} & 0.191^{*} & 0.374^{***} \\ 
  & (0.063) & (0.062) & (0.069) & (0.063) & (0.056) & (0.067) & (0.054) & (0.053) & (0.051) & (0.075) & (0.072) \\ 
  Sat & 0.265^{***} & 0.243^{***} & 0.248^{***} & 0.276^{***} & 0.293^{***} & 0.277^{***} & 0.292^{***} & 0.223^{***} & 0.214^{***} & 0.135^{***} & 0.162^{***} \\ 
  & (0.024) & (0.024) & (0.028) & (0.026) & (0.022) & (0.026) & (0.021) & (0.021) & (0.020) & (0.035) & (0.036) \\ 
  Sun & -1.291^{***} & -1.355^{***} & -1.321^{***} & -1.213^{***} & -1.140^{***} & -1.261^{***} & -1.173^{***} & -1.203^{***} & -1.291^{***} & -1.398^{***} & -1.514^{***} \\ 
  & (0.028) & (0.027) & (0.031) & (0.029) & (0.026) & (0.029) & (0.025) & (0.025) & (0.022) & (0.037) & (0.036) \\ 
  Mon & -0.135^{***} & -0.188^{***} & -0.163^{***} & -0.175^{***} & -0.010 & -0.062^{*} & -0.037 & -0.093^{***} & -0.123^{***} & -0.139^{***} & -0.222^{***} \\ 
  & (0.028) & (0.028) & (0.032) & (0.030) & (0.026) & (0.029) & (0.026) & (0.025) & (0.022) & (0.035) & (0.036) \\ 
  Tue & -0.225^{***} & -0.263^{***} & -0.271^{***} & -0.251^{***} & -0.170^{***} & -0.233^{***} & -0.207^{***} & -0.200^{***} & -0.226^{***} & -0.229^{***} & -0.305^{***} \\ 
  & (0.028) & (0.028) & (0.032) & (0.030) & (0.026) & (0.029) & (0.026) & (0.025) & (0.022) & (0.035) & (0.036) \\ 
  Wed & -0.247^{***} & -0.250^{***} & -0.266^{***} & -0.271^{***} & -0.209^{***} & -0.243^{***} & -0.232^{***} & -0.249^{***} & -0.266^{***} & -0.262^{***} & -0.312^{***} \\ 
  & (0.027) & (0.027) & (0.030) & (0.028) & (0.025) & (0.028) & (0.025) & (0.025) & (0.021) & (0.037) & (0.036) \\ 
  Thr & -0.218^{***} & -0.218^{***} & -0.207^{***} & -0.239^{***} & -0.201^{***} & -0.211^{***} & -0.210^{***} & -0.234^{***} & -0.230^{***} & -0.249^{***} & -0.258^{***} \\ 
  & (0.024) & (0.024) & (0.028) & (0.026) & (0.022) & (0.026) & (0.021) & (0.021) & (0.020) & (0.035) & (0.036) \\  
 \hline \\[-1.8ex] 
Observations & \multicolumn{1}{c}{386} & \multicolumn{1}{c}{386} & \multicolumn{1}{c}{386} & \multicolumn{1}{c}{386} & \multicolumn{1}{c}{386} & \multicolumn{1}{c}{386} & \multicolumn{1}{c}{386} & \multicolumn{1}{c}{386} & \multicolumn{1}{c}{386} & \multicolumn{1}{c}{386} & \multicolumn{1}{c}{386} \\ 
$\sigma^{2}$ & \multicolumn{1}{c}{0.022} & \multicolumn{1}{c}{0.022} & \multicolumn{1}{c}{0.022} & \multicolumn{1}{c}{0.017} & \multicolumn{1}{c}{0.014} & \multicolumn{1}{c}{0.022} & \multicolumn{1}{c}{0.014} & \multicolumn{1}{c}{0.011} & \multicolumn{1}{c}{0.013} & \multicolumn{1}{c}{0.027} & \multicolumn{1}{c}{0.017} \\ 
Akaike Inf. Crit. & \multicolumn{1}{c}{-355.744} & \multicolumn{1}{c}{-358.769} & \multicolumn{1}{c}{-352.505} & \multicolumn{1}{c}{-453.681} & \multicolumn{1}{c}{-519.897} & \multicolumn{1}{c}{-366.579} & \multicolumn{1}{c}{-539.814} & \multicolumn{1}{c}{-620.166} & \multicolumn{1}{c}{-577.969} & \multicolumn{1}{c}{-282.397} & \multicolumn{1}{c}{-460.103} \\ 
\hline 
\hline \\[-1.8ex] 
\textit{Note:}  & \multicolumn{11}{r}{$^{\boldsymbol{\cdot}}$p$<$0.1; $^{*}$p$<$0.05; $^{**}$p$<$0.01; $^{***}$p$<$0.001} \\ 
\end{tabular}
 
}
\end{sidewaystable}

\begin{figure}[h!]
\caption{Daily time series of unit sold, price per unit and autocorrelation function for the $11$ store brands. The red vertical bar indicates the intervention date.}
\label{fig:tp_des}
\centering
\includegraphics[scale=0.63]{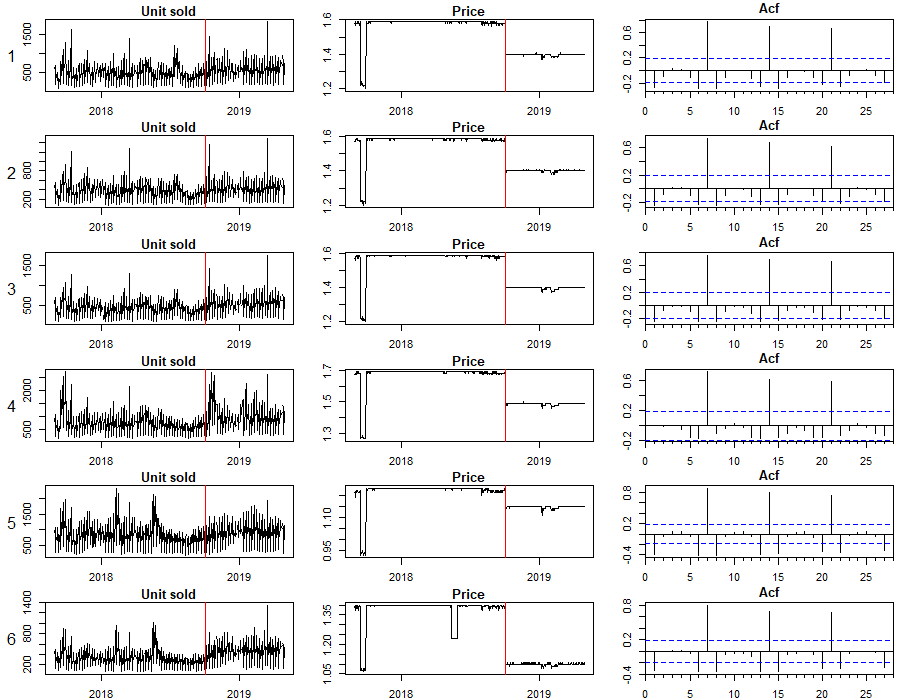}
\includegraphics[scale=0.63]{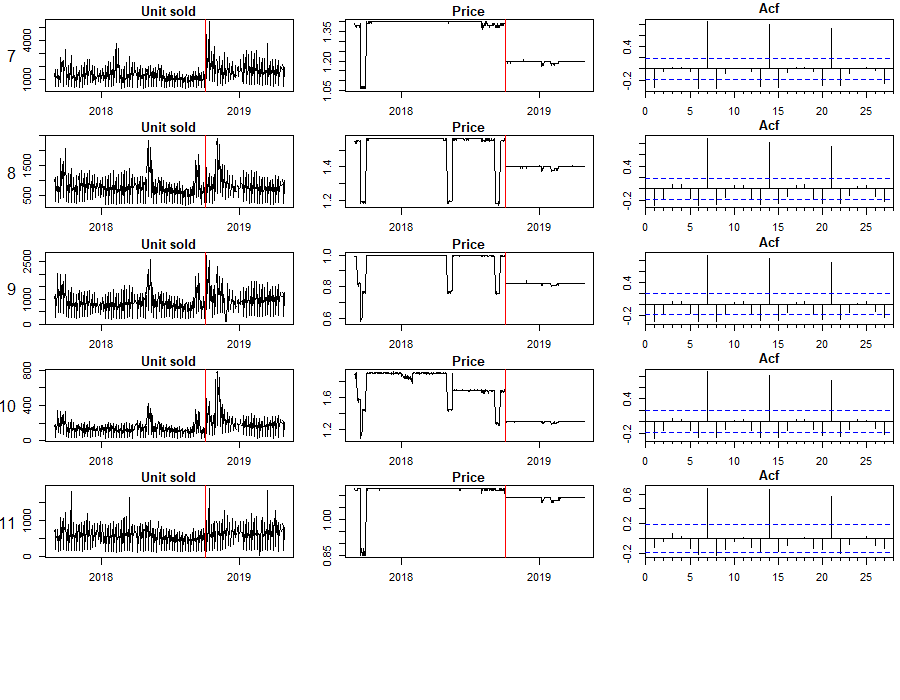}
\end{figure}

\begin{figure}[h!]
\caption{Residual diagnostics (autocorrelation functions and Normal QQ plots) of the C-ARIMA models fitted to the time series of units sold (in log scale).}
\label{fig:tp_arima_res}
\centering
\begin{tabular}{cc}
\includegraphics[scale=0.4]{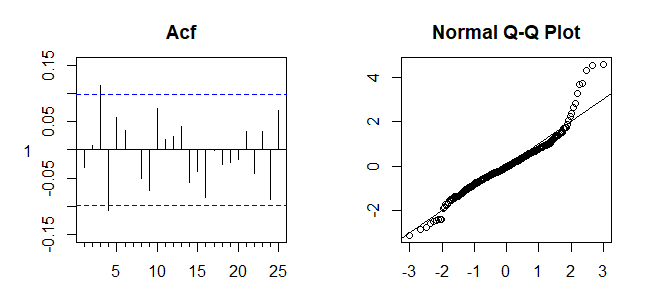} & \includegraphics[scale=0.4]{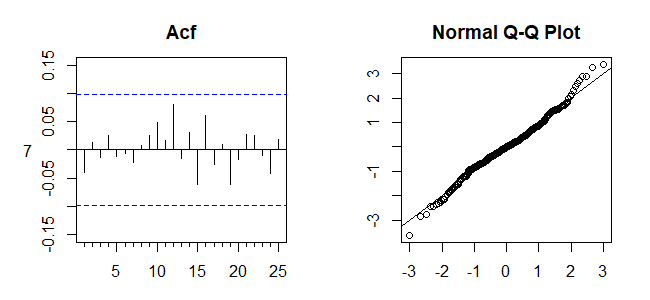} \\
\includegraphics[scale=0.4]{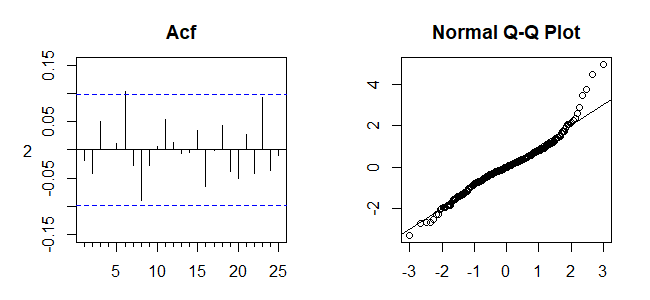} & \includegraphics[scale=0.4]{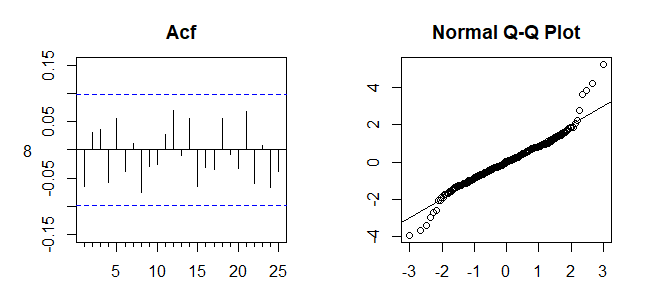} \\
\includegraphics[scale=0.4]{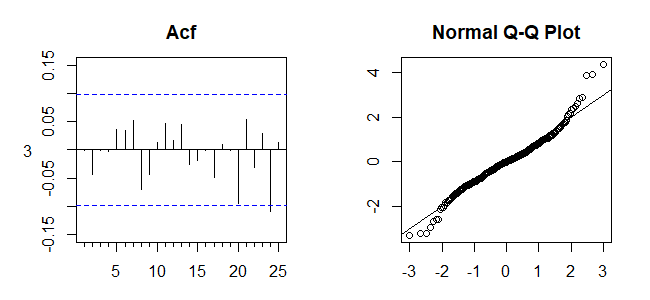} & \includegraphics[scale=0.4]{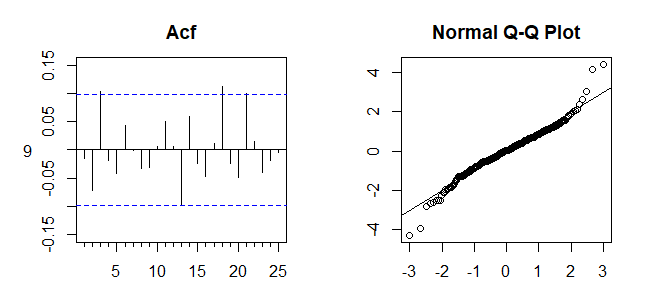} \\
\includegraphics[scale=0.4]{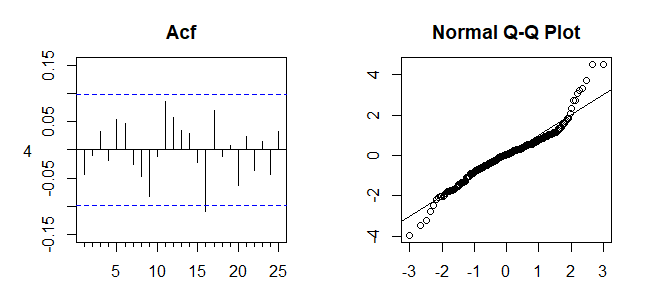} & \includegraphics[scale=0.4]{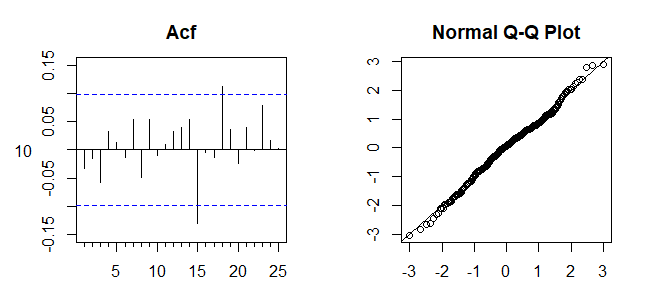} \\
\includegraphics[scale=0.4]{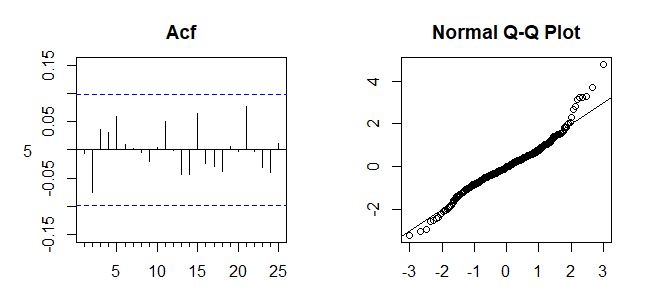} & \includegraphics[scale=0.4]{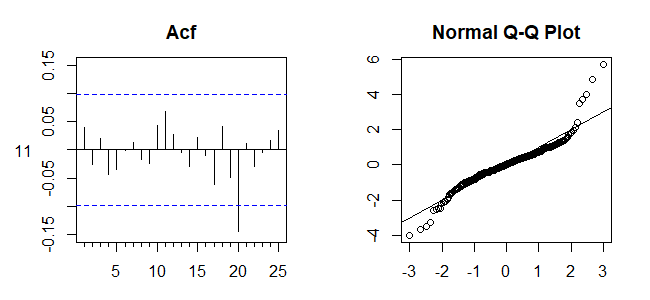} \\
\includegraphics[scale=0.4]{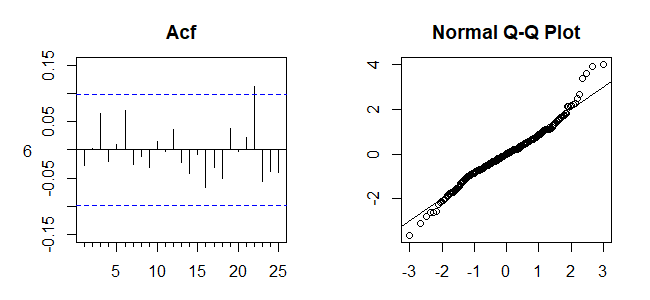} &  \\
\end{tabular}
\end{figure}

\begin{figure}
\centering
\caption{Point causal effect of the new price policy on the sales of store-brand products for each time horizon ($1$, $3$ and $6$ months) estimated via C-ARIMA (the dependent variable is the daily sales counts of each product in log scale).}
\label{fig:tp_arima_causal}
\includegraphics[scale=0.53]{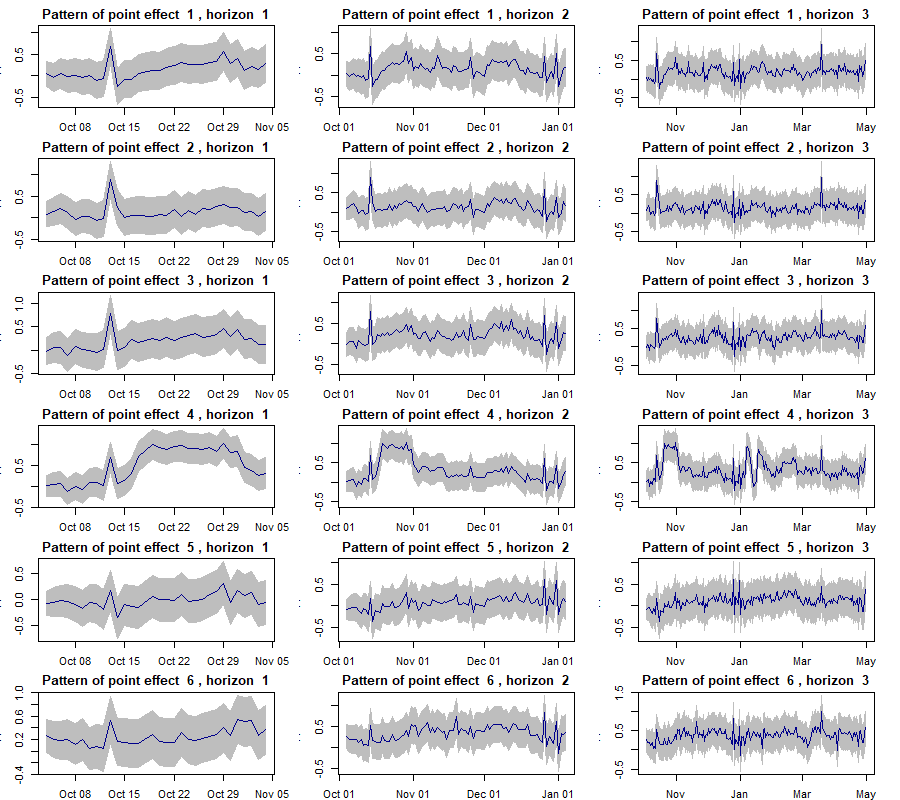}
\includegraphics[scale=0.53]{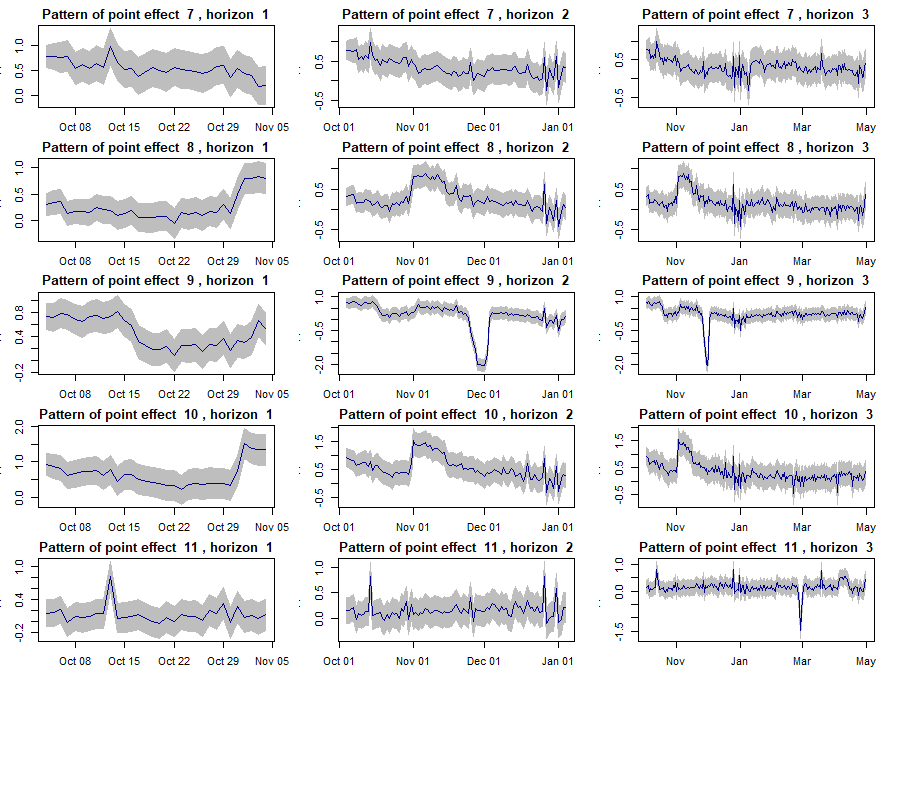}
\end{figure}

\begin{figure}
\centering
\caption{Observed sales (grey) and forecasted sales (blue) of each store brand and for each time horizon ($1$, $3$ and $6$ months). The red vertical bar indicates the intervention date.}
\label{fig:tp_arima_forecast}
\includegraphics[scale=0.53]{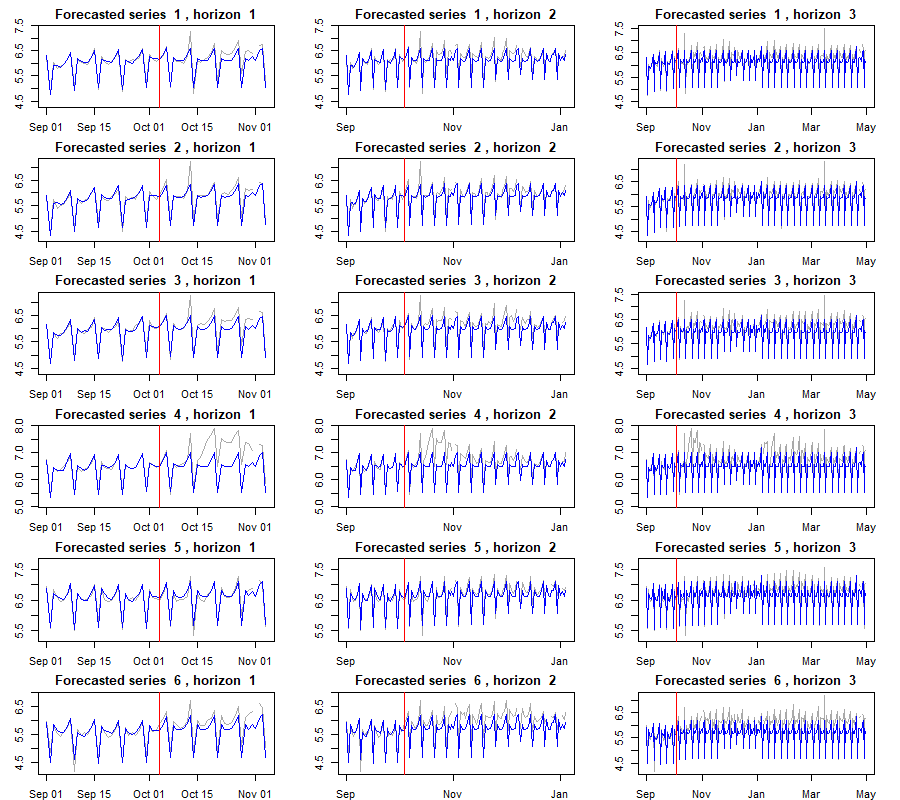}
\includegraphics[scale=0.53]{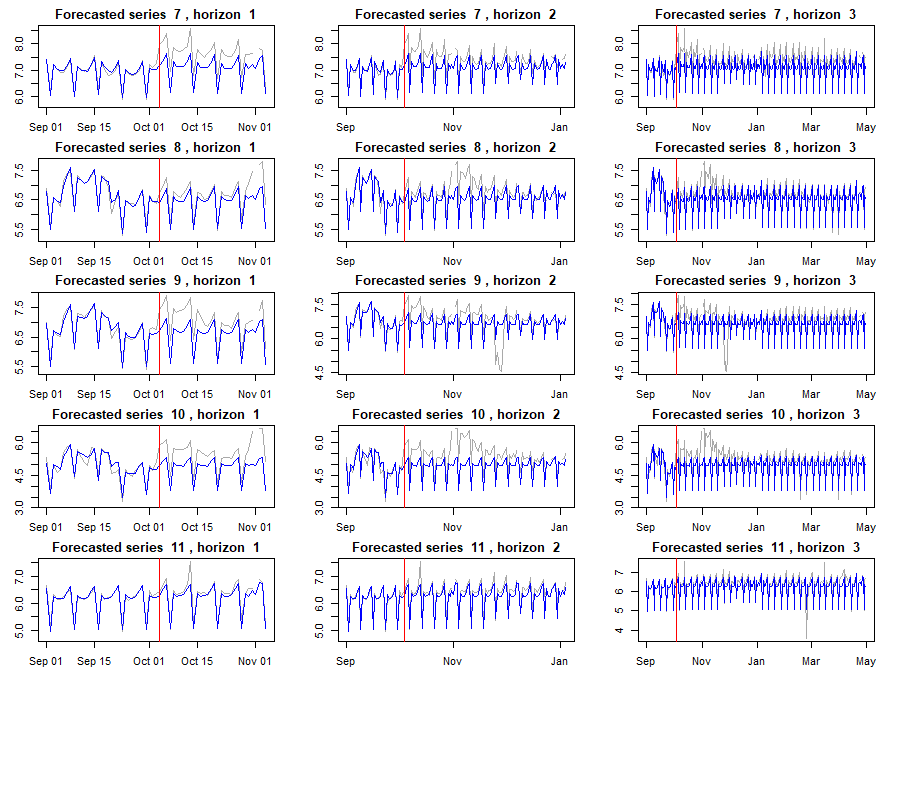}
\end{figure}

\clearpage
\subsection{Competitor brands}
\label{appA_competitor}

\begin{table}[h!]
    \centering
    \caption{Causal effect estimates of the permanent price rebate on sales of competitor-brand cookies after one month, three months and six months from the intervention. In this table, $ \hat{\bar{\tau}}^Y_t$ is the estimated temporal average effect on the original variable ($ \hat{\bar{\tau}}^Y_t = 0 $ implies no effect) and the empirical critical values are computed based on Equation (\ref{eqn:avetauhat-H0_2}) by bootstrapping the errors from the model residuals.}
    \label{tab:boot_comp}
    \scalebox{0.8}{
    \begin{tabular}{l D{.}{.}{-2} r D{.}{.}{-2} r D{.}{.}{-2}} 
\\[-1.8ex]\hline 
\hline \\[-1.8ex] 
      & \multicolumn{5}{c}{\textit{Time horizon:}} \\ 
\cline{2-6}
\\[-1.8ex]     & $1 \text{ month}$  &  & $3 \text{ months}$  &  & $6 \text{ months}$ \\                      
Item  &  \hat{\bar{\tau}}_t   &  & \hat{\bar{\tau}}_t  &  & \hat{\bar{\tau}}_t  \\ \hline \\[-1.8ex] 
 1    &  -0.04 & &  0.02      & & 0.04 \\
 2    &  -0.13 & & -0.07      & & -0.15   \\
 3    &  0.04  & & 0.09       & & 0.17 \\
 4    &  0.00  & & -0.13      & & -0.04  \\
 5    &  -0.03 & & 0.05       & & 0.12^{**}     \\
 6    &  -0.05 & & 0.03       & & 0.09  \\
 7    &  0.04  & & 0.11       & & 0.4^{.}  \\
 8    &  -0.09 & & -0.06      & & -0.08^{*}   \\
 9    &  -0.09 & & -0.11      & & -0.1  \\
10    &  -0.03 & & -0.12^{**} & & -0.11^{***}  \\
\hline
\hline \\[-1.8ex]
\textit{Note:}      & \multicolumn{5}{r}{$^{\boldsymbol{\cdot}}$p$<$0.1; $^{*}$p$<$0.05; $^{**}$p$<$0.01; $^{***}$p$<$0.001} \\                  
\end{tabular}
    }
\end{table}

\begin{sidewaystable}[h!]
\caption{Estimated coefficients (standard errors within parentheses) of the C-ARIMA models fitted to the $10$ competitor brands in the pre-intervention period. The dependent variable is the daily sales counts of each product in log scale.}
\scalebox{0.55}{
\begin{tabular}{@{\extracolsep{5pt}}lD{.}{.}{-3} D{.}{.}{-3} D{.}{.}{-3} D{.}{.}{-3} D{.}{.}{-3} D{.}{.}{-3} D{.}{.}{-3} D{.}{.}{-3} D{.}{.}{-3} D{.}{.}{-3} } 
\\[-1.8ex]\hline 
\hline \\[-1.8ex] 
 & \multicolumn{10}{c}{\textit{Dependent variable:}} \\ 
\cline{2-11} 
\\[-1.8ex] & \multicolumn{1}{c}{Item 1} & \multicolumn{1}{c}{Item 2} & \multicolumn{1}{c}{Item 3} & \multicolumn{1}{c}{Item 4} & \multicolumn{1}{c}{Item 5} & \multicolumn{1}{c}{Item 6} & \multicolumn{1}{c}{Item 7} & \multicolumn{1}{c}{Item 8} & \multicolumn{1}{c}{Item 9} & \multicolumn{1}{c}{Item 10} \\ 
\hline \\[-1.8ex] 
  $\phi_1$ &  0.947^{***} & 0.959^{***} & 0.949^{***} & -0.090^{*} & 0.876^{***} & 0.973^{***} & 1.800^{***} & 0.939^{***} & 0.855^{***} & 0.786^{***} \\ 
  & (0.015) & (0.014) & (0.016) & (0.043) & (0.053) & (0.157) & (0.088) & (0.028) & (0.034) & (0.083) \\ 
  $\phi_2$ &  &  &  & 0.850^{***} & -0.161^{**} & -0.075 & -0.821^{***} &  &  &  \\ 
  &  &  &  & (0.038) & (0.054) & (0.130) & (0.081) &  &  &  \\ 
  $\phi_3$ &  &  &  &  &  &  &  &  &  &  \\ 
  &  &  &  &  &  &  &  &  &  &  \\ 
  $\theta_1$ &  &  &  & 0.940^{***} &  & -0.408^{**} & -0.768^{***} & -0.566^{***} & -0.239^{***} & -0.514^{***} \\ 
  &  &  &  & (0.047) &  & (0.142) & (0.109) & (0.068) & (0.068) & (0.120) \\ 
  $\theta_2$ &  &  &  &  &  &  &  & -0.117^{.} &  &  \\ 
  &  &  &  &  &  &  &  & (0.070) &  &  \\ 
  $\Phi_1$ &  & 0.108^{*} & 0.907^{***} & 0.224^{***} & 0.162^{**} & 0.852^{***} &  &  & 0.281^{***} &  \\ 
  &  & (0.054) & (0.030) & (0.054) & (0.053) & (0.088) &  &  & (0.053) &  \\ 
  $\Theta_1$ &  &  & -1.000^{***} &  &  & -0.779^{***} &  &  &  &  \\ 
  &  &  & (0.021) &  &  & (0.103) &  &  &  &  \\ 
  c & 8.296^{***} & 7.862^{***} & 7.828^{***} & 8.335^{***} & 8.873^{***} & 8.380^{***} & 10.189^{***} & 7.029^{***} & 6.773^{***} & 6.327^{***} \\ 
  & (0.300) & (0.297) & (0.150) & (0.162) & (0.137) & (0.127) & (0.292) & (0.048) & (0.050) & (0.108) \\ 
  price & -2.465^{***} & -2.350^{***} & -2.074^{***} & -2.193^{***} & -2.210^{***} & -2.257^{***} & -3.742^{***} & -1.490^{***} & -1.308^{***} & -2.499^{***} \\ 
  & (0.164) & (0.124) & (0.135) & (0.130) & (0.139) & (0.121) & (0.286) & (0.089) & (0.127) & (0.220) \\ 
  hol & 0.157^{***} & 0.161^{***} & 0.099^{*} & 0.154^{***} & 0.107^{**} & 0.155^{***} & 0.138^{*} & 0.178^{***} & 0.131^{***} & 0.165^{***} \\ 
  & (0.059) & (0.044) & (0.048) & (0.045) & (0.033) & (0.029) & (0.062) & (0.025) & (0.029) & (0.038) \\ 
  Dec.Sun & 0.404^{***} & 0.538^{***} & 0.446^{***} & 0.515^{***} & 0.431^{***} & 0.379^{***} & 0.363^{***} & 0.413^{***} & 0.338^{***} & 0.437^{***} \\ 
  & (0.092) & (0.076) & (0.068) & (0.085) & (0.057) & (0.056) & (0.095) & (0.046) & (0.063) & (0.069) \\ 
  Sat & 0.323^{***} & 0.262^{***} & 0.272^{***} & 0.244^{***} & 0.322^{***} & 0.357^{***} & 0.315^{***} & 0.216^{***} & 0.263^{***} & 0.215^{***} \\ 
  & (0.036) & (0.030) & (0.013) & (0.034) & (0.024) & (0.027) & (0.037) & (0.018) & (0.025) & (0.027) \\ 
  Sun & -0.982^{***} & -1.143^{***} & -1.083^{***} & -1.351^{***} & -1.011^{***} & -0.992^{***} & -1.057^{***} & -1.303^{***} & -1.288^{***} & -1.303^{***} \\ 
  & (0.047) & (0.039) & (0.018) & (0.045) & (0.034) & (0.031) & (0.050) & (0.020) & (0.030) & (0.029) \\ 
  Mon & -0.104^{*} & -0.182^{***} & -0.172^{***} & -0.199^{***} & -0.050 & -0.085^{**} & 0.017 & -0.117^{***} & -0.119^{***} & -0.119^{***} \\ 
  & (0.051) & (0.042) & (0.018) & (0.048) & (0.037) & (0.032) & (0.054) & (0.020) & (0.031) & (0.029) \\ 
  Tue & -0.254^{***} & -0.240^{***} & -0.266^{***} & -0.267^{***} & -0.206^{***} & -0.241^{***} & -0.166^{**} & -0.208^{***} & -0.231^{***} & -0.229^{***} \\ 
  & (0.051) & (0.042) & (0.018) & (0.048) & (0.037) & (0.032) & (0.054) & (0.020) & (0.031) & (0.029) \\ 
  Wed & -0.237^{***} & -0.222^{***} & -0.284^{***} & -0.285^{***} & -0.197^{***} & -0.254^{***} & -0.218^{***} & -0.237^{***} & -0.242^{***} & -0.270^{***} \\ 
  & (0.046) & (0.039) & (0.016) & (0.044) & (0.033) & (0.030) & (0.049) & (0.020) & (0.029) & (0.028) \\ 
  Thr & -0.208^{***} & -0.182^{***} & -0.220^{***} & -0.237^{***} & -0.187^{***} & -0.232^{***} & -0.200^{***} & -0.196^{***} & -0.219^{***} & -0.234^{***} \\ 
  & (0.036) & (0.030) & (0.012) & (0.034) & (0.024) & (0.027) & (0.037) & (0.018) & (0.025) & (0.027) \\ 
 \hline \\[-1.8ex] 
Observations & \multicolumn{1}{c}{386} & \multicolumn{1}{c}{386} & \multicolumn{1}{c}{385} & \multicolumn{1}{c}{386} & \multicolumn{1}{c}{386} & \multicolumn{1}{c}{386} & \multicolumn{1}{c}{386} & \multicolumn{1}{c}{386} & \multicolumn{1}{c}{386} & \multicolumn{1}{c}{386} \\ 
$\sigma^{2}$ & \multicolumn{1}{c}{0.081} & \multicolumn{1}{c}{0.046} & \multicolumn{1}{c}{0.049} & \multicolumn{1}{c}{0.044} & \multicolumn{1}{c}{0.021} & \multicolumn{1}{c}{0.017} & \multicolumn{1}{c}{0.094} & \multicolumn{1}{c}{0.013} & \multicolumn{1}{c}{0.017} & \multicolumn{1}{c}{0.027} \\ 
Akaike Inf. Crit. & \multicolumn{1}{c}{147.413} & \multicolumn{1}{c}{-66.040} & \multicolumn{1}{c}{-33.990} & \multicolumn{1}{c}{-85.448} & \multicolumn{1}{c}{-372.686} & \multicolumn{1}{c}{-451.392} & \multicolumn{1}{c}{207.605} & \multicolumn{1}{c}{-575.002} & \multicolumn{1}{c}{-449.421} & \multicolumn{1}{c}{-281.026} \\ 
\hline 
\hline \\[-1.8ex] 
\textit{Note:}  & \multicolumn{10}{r}{$^{\boldsymbol{\cdot}}$p$<$0.1; $^{*}$p$<$0.05; $^{**}$p$<$0.01; $^{***}$p$<$0.001} \\ 
\end{tabular} 
}
\end{sidewaystable}

\begin{figure}[h!]
\caption{Daily time series of unit sold, price per unit, relative price to store brand and autocorrelation function for the $10$ competitor brands. The red vertical bar indicates the intervention date.}
\label{fig:sp_des}
\centering
\includegraphics[scale=0.65]{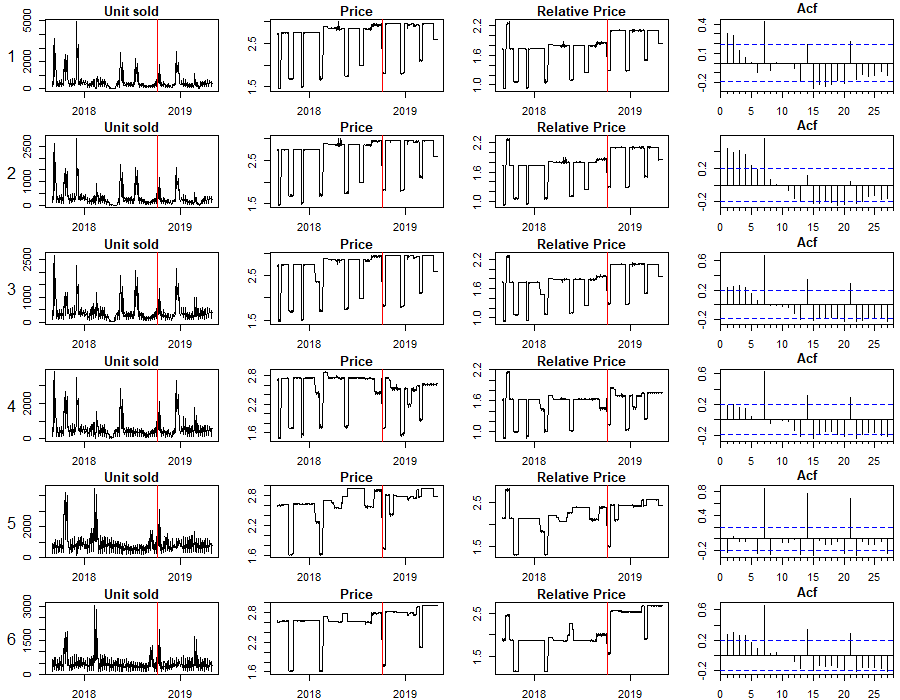}
\includegraphics[scale=0.65]{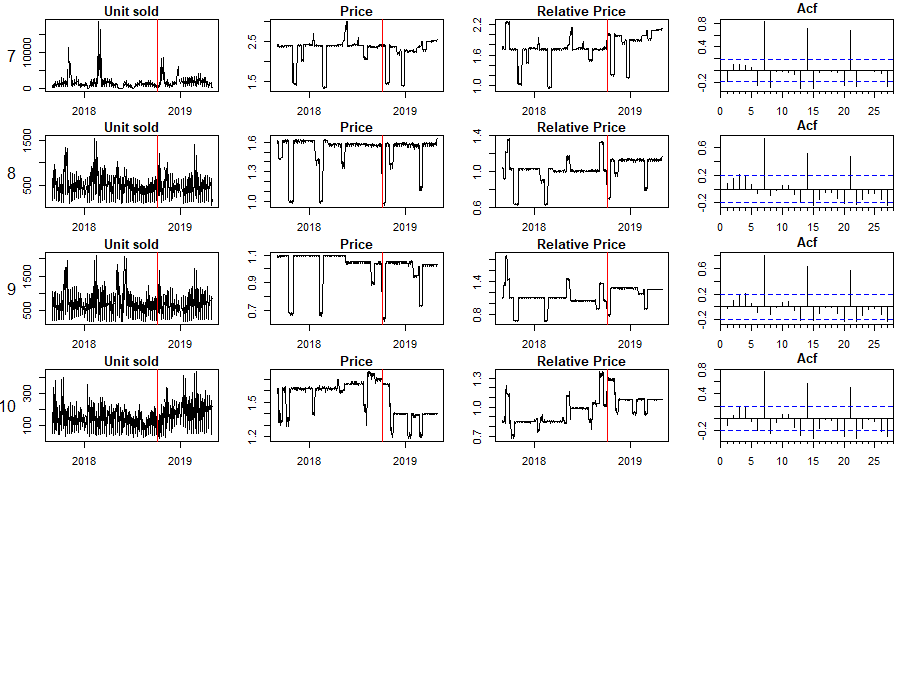}
\end{figure}

\begin{figure}
\caption{Residual diagnostics (autocorrelation functions and Normal QQ plots) of the C-ARIMA models fitted to the time series of units sold (in log scale).}
\label{fig:sp_arima_res}
\centering
\begin{tabular}{cc}
\includegraphics[scale=0.5]{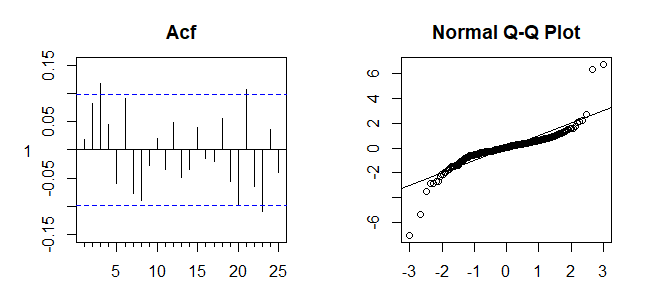} & \includegraphics[scale=0.5]{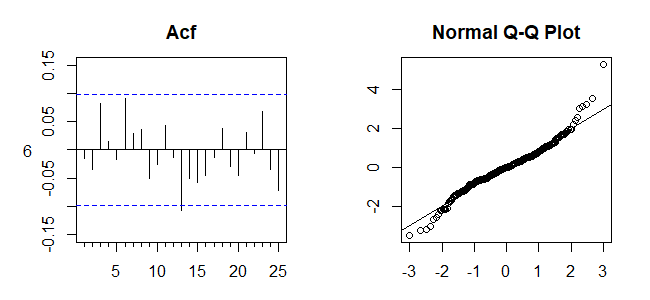} \\
\includegraphics[scale=0.5]{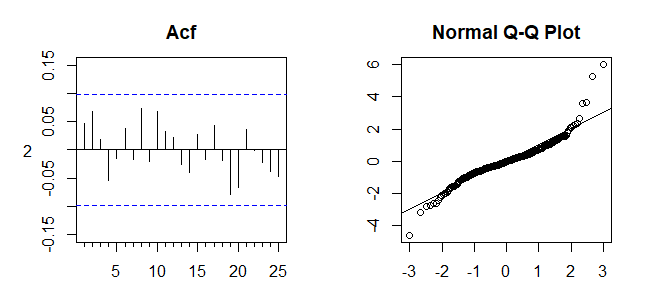} & \includegraphics[scale=0.5]{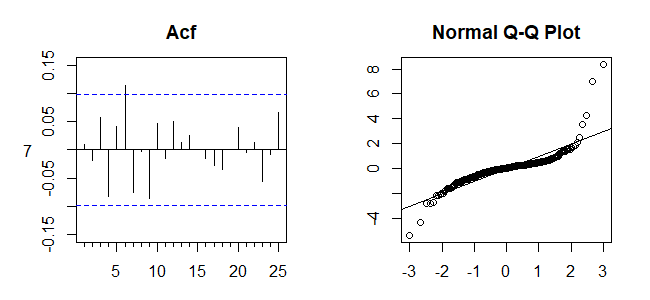} \\
\includegraphics[scale=0.5]{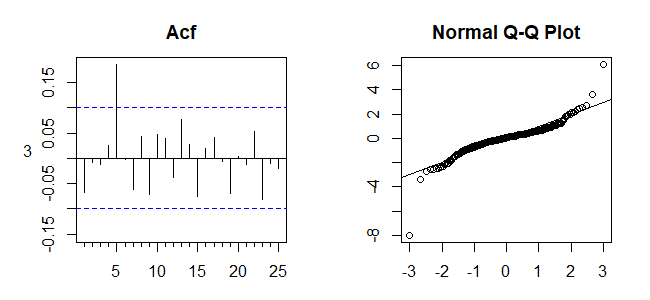} & \includegraphics[scale=0.5]{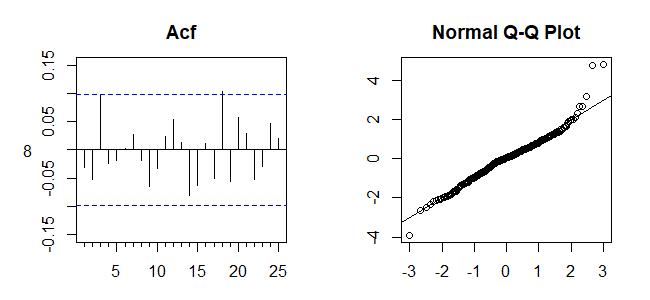} \\
\includegraphics[scale=0.5]{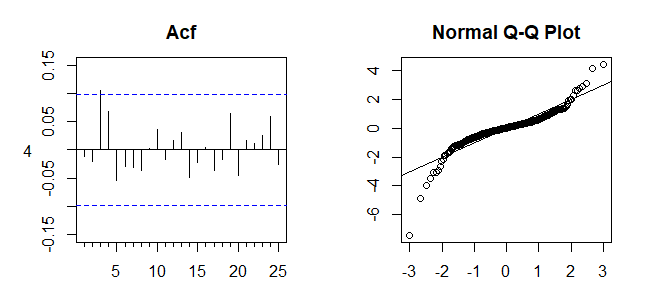} & \includegraphics[scale=0.5]{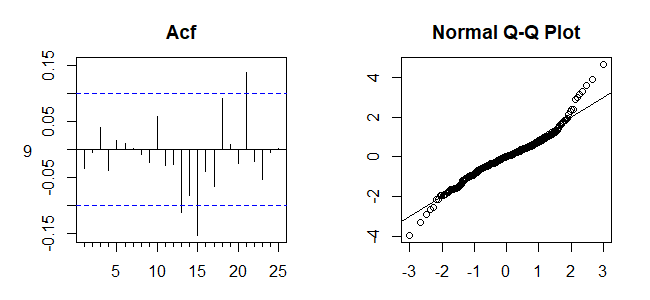} \\
\includegraphics[scale=0.5]{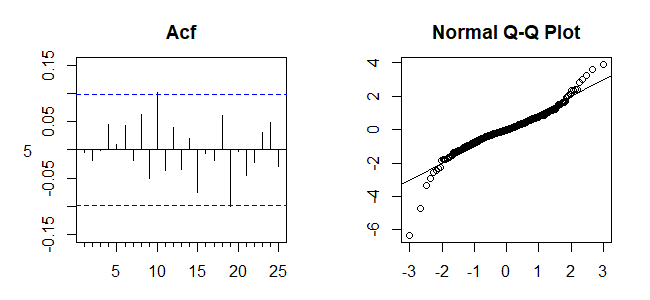} & \includegraphics[scale=0.5]{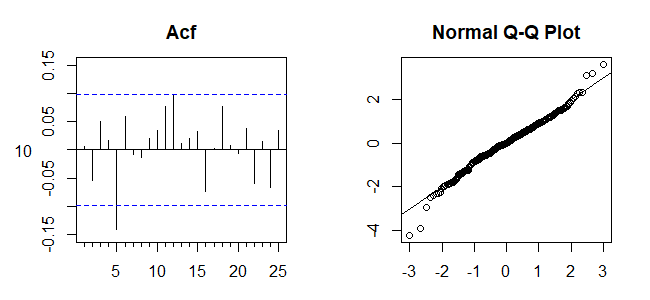} \\
\end{tabular}
\end{figure}

\begin{figure}
\centering
\caption{Point causal effect of the new price policy on the sales of competitor-brand products for each time horizon ($1$, $3$ and $6$ months) estimated via C-ARIMA (the dependent variable is the daily sales counts of each product in log scale).}
\label{fig:sp_arima_effect}
\includegraphics[scale=0.53]{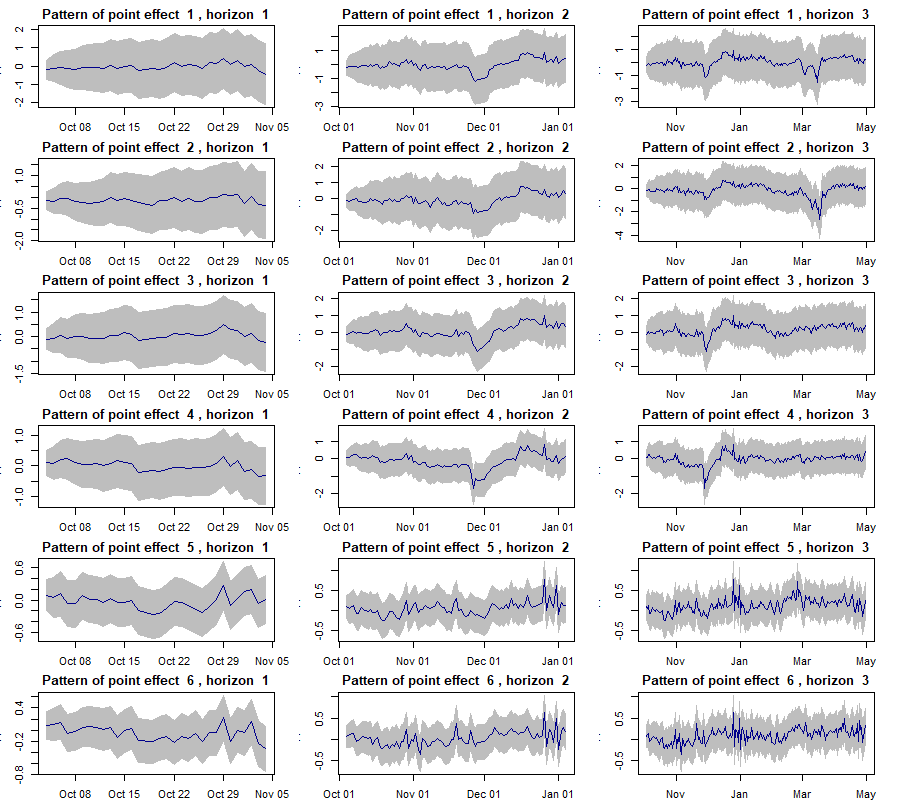}
\includegraphics[scale=0.53]{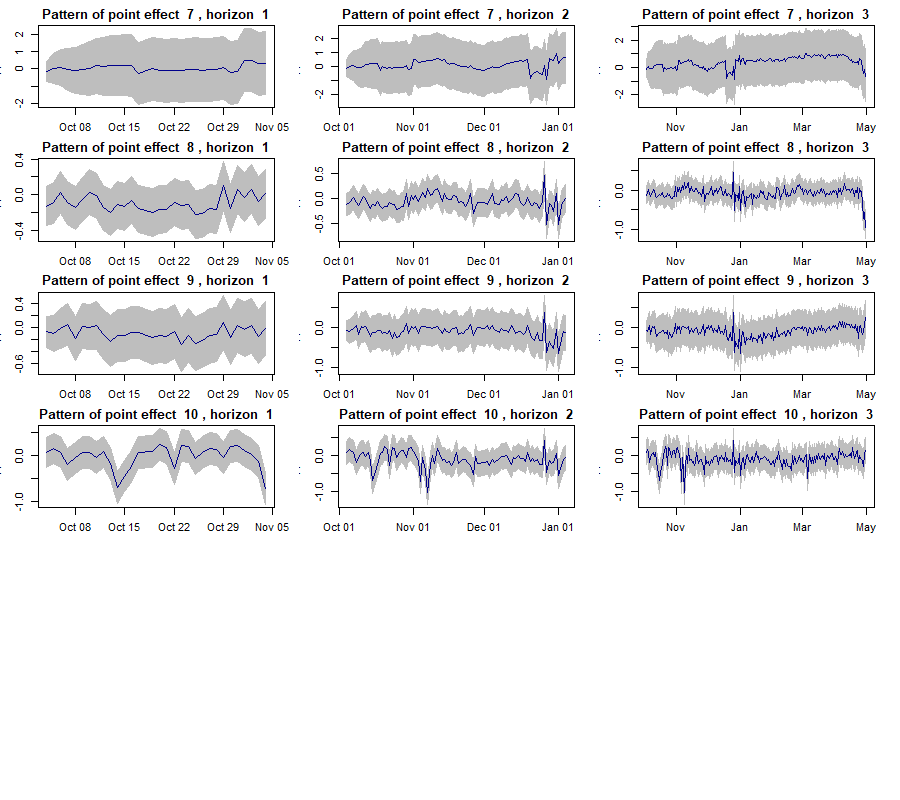}
\end{figure}

\begin{figure}
\centering
\caption{Observed sales (grey) and forecasted sales (blue) of each competitor brand and for each time horizon ($1$, $3$ and $6$ months). The red vertical bar indicates the intervention date.}
\label{fig:sp_arima_forecast}
\includegraphics[scale=0.53]{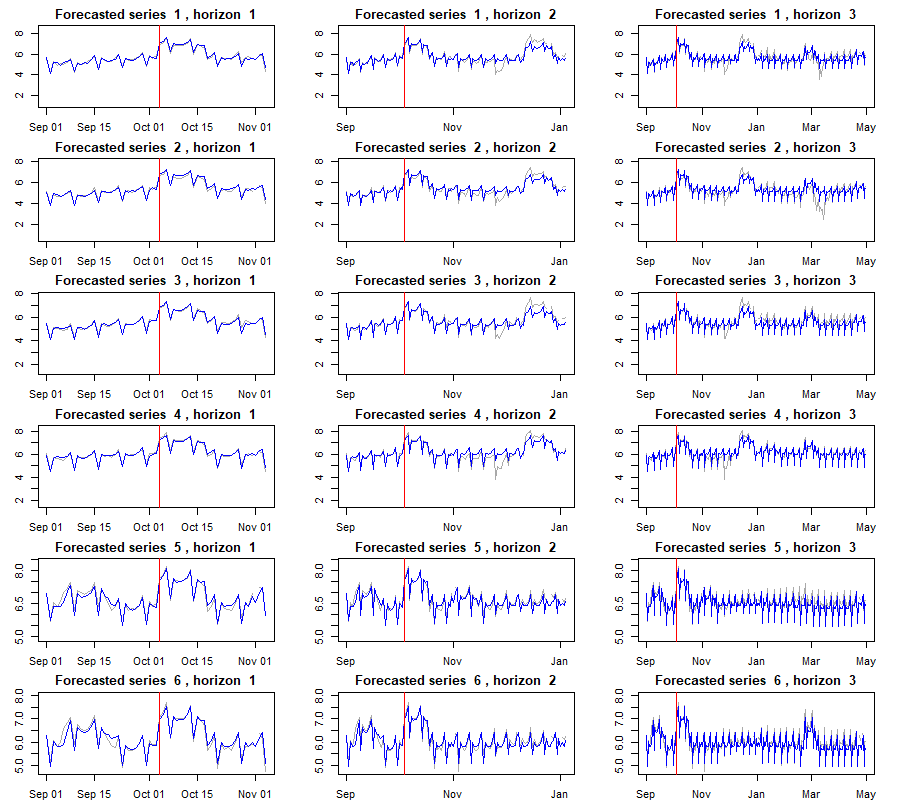}
\includegraphics[scale=0.53]{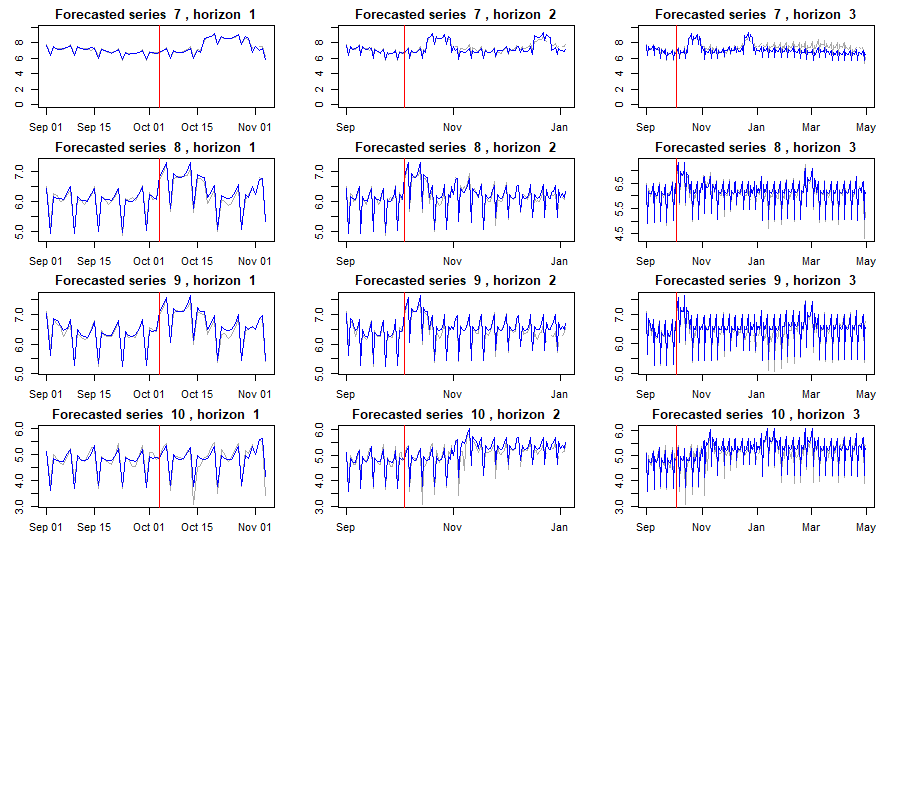}
\end{figure}

\clearpage
\section{}
\label{appB}
\numberwithin{equation}{section}
\setcounter{equation}{0}

\subsection{Proof of Theorem \ref{theo}} 
\label{appB_proof}
Considering the setup formalized in Section \ref{sect:cARIMA}, we can write the estimator of the point effect (Equation~(\ref{eqn:tauhat})) as,
\begin{equation}
\label{eqn:error}
\hat{\tau}_{t^*+k}(1;0) =  
z_{t^* + k} + \tau_{t^* + k}(1;0) - \widehat{z}_{t^* + k | t^*} = 
\sum_{i = 0}^{k-1} \psi_{i} \varepsilon_{t^* + k - i} + \tau_{t^* + k}(1;0),
\end{equation}
where the last expression comes from the well known relationship that the forecasting error at lag $k$ is a zero-mean $\MA(k-1)$. Therefore, under the null hypothesis that the intervention has no effect, namely $H_0: \tau_t(1;0) = 0$ for all $t \in \{t^*+1, \dots, T \}$, Equation (\ref{eqn:error}) implies that the estimators of the point, cumulative, and temporal average effects (Equations~(\ref{eqn:tauhat})-(\ref{eqn:avetauhat})) can be arranged as
\begin{equation*}
  \hat{\tau}_{t^*+k}(1;0) | H_0 = \sum_{i = 0}^{k-1} \psi_{i} \varepsilon_{t^* + k - i}
\end{equation*}
\begin{equation*}
  \hat{\Delta}_{t^*+k}(1;0) 
  |H_0 = 
  \sum_{h = 1}^k \varepsilon_{t^*+h} \sum_{i = 0}^{k-h} \psi_i
\end{equation*}
\begin{equation*}
  \hat{\bar{\tau}}_{t^*+k}(1;0)  
  |H_0 = 
  \frac{1}{k} \sum_{h = 1}^k \varepsilon_{t^*+h} \sum_{i = 0}^{k-h} \psi_i.
\end{equation*}
In case the $\varepsilon_{t}$ error term is normally distributed, above equations become
\begin{equation*}
  \hat{\tau}_{t^*+k}(1;0)|H_0 
\sim{} N
\left[ 
0,
\sigma^{2}_{\varepsilon} \sum_{h = 0}^{k-1} \psi_{h}^{2}	
\right]
\end{equation*}
\begin{equation*}
\hat{\Delta}_{t^*+k}(1;0)|H_0 
\sim{} N
\left[ 
0,
\sigma^{2}_{\varepsilon} \sum_{h = 1}^{k} \left( \sum_{i = 0}^{k - h} \psi_{i} \right)^{2}	
\right]
\end{equation*}
\begin{equation*}
\hat{\bar{\tau}}_{t^*+k}(1;0)|H_0 
\sim{}
\left[ 
0,
\frac{\sigma^{2}_{\varepsilon}}{k^2} \sum_{h = 1}^{k} \left( \sum_{i = 0}^{k - h} \psi_{i} \right)^{2}	
\right].
\end{equation*}

\newpage

\subsection{Proof of Theorem \ref{theo_2}}
\label{appB_untransformed}

The derivation of Theorem \ref{theo_2} follows directly from Theorem \ref{theo}. Indeed, the transformation to achieve stationarity can also be expressed as $(1-L^S)^D (1-L)^d = \sum\limits_{j = 0}^A a_j L^j$ for some $a_j$ coefficients with $a_0 = 1$ and $A = SD +d $; then we have,
\begin{equation}
  \label{eqn:tauh-11}
   \hat{\tau}_{t^*+k}(1;0) =  \sum\limits_{j = 0}^A a_j L^j \hat{\tau}^Y_{t^*+k}(1;0).  
\end{equation}
Standard algebra of polynomial convolution leads in general to invert (\ref{eqn:tauh-11}) as   
\begin{equation}
  \label{eqn:tauhY-11}
   \hat{\tau}_{t^*+k}^Y (1;0)
   = \sum\limits_{j = 0}^\infty b_j L^j \hat{\tau}_{t^*+k}(1;0)
   = \sum\limits_{j = 0}^\infty b_j \hat{\tau}_{t^*+k-j}(1;0)
\end{equation}
where the $b_j$ coefficients are recursively computed from the $a_j$'s via 
\begin{align*}
  b_j = 
  \left\{
  \begin{array}{cc}
     1                                                     & \text{ if } j = 0\\
    - \displaystyle \sum_{i = 1}^{\min\{A; j\}} a_i b_{j-i}  & \text{ if } j > 0
  \end{array}
  \right..
\end{align*}
On the other hand, $\hat{\tau}_{t^*+k}(1;0) = 0$ for $k \leq 0$, so that (\ref{eqn:tauhY-11}) becomes
\begin{equation}
  \hat{\tau}_{t^*+k}^Y (1;0) = \sum\limits_{j = 0}^{k-1} b_j \hat{\tau}_{t^*+k-j} (1;0).
\end{equation}

From Equation (\ref{eqn:tauhY-11}) and Theorem \ref{theo} it follows,

$$\hat{\tau}^Y_{t^*+k} (1;0) = \sum_{j = 0}^{k-1} b_j \hat{\tau}_{t^*+k-j} (1;0) = \sum_{j = 0}^{k-1} b_j \sum_{i =0}^{k-j-1} \psi_i \varepsilon_{t^*+k-j-i}.$$

From the above equation we can derive the following estimators for the point, cumulative and temporal average effects on the original variable, 
\begin{equation*}
  \hat{\tau}_{t^*+k}^{Y}(1;0) | H_0 = \sum_{j = 1}^{k} \varepsilon_{t^*+j} \sum\limits_{i = 0}^{k-j} b_i \psi_{k-j-i}
\end{equation*}

\begin{equation*}
    \hat{\Delta}_{t^*+k}^{Y}(1;0) | H_0 = \sum_{h = 1}^k \varepsilon_{t^*+h} \sum_{i = 0}^{k-h} b_i \sum_{j = i}^{k-h} \psi_{k-h-j}
\end{equation*}

\begin{equation*}
    \hat{\bar{\tau}}_{t^*+k}^{Y}(1;0) | H_0 = \frac{1}{k}\sum_{h = 1}^k \varepsilon_{t^*+h} \sum_{i = 0}^{k-h} b_i \sum_{j = i}^{k-h} \psi_{k-h-j}
\end{equation*}

Assuming normality of the error term,

\begin{equation*}
  \hat{\tau}_{t^*+k}^{Y}(1;0) | H_0 \sim N \left[ 0, \sigma^2_{\varepsilon} \sum_{j=1}^k \left( \sum_{i = 0}^{k-j} b_i \psi_{k-j-i} \right)^2 \right]
\end{equation*}

\begin{equation*}
    \hat{\Delta}_{t^*+k}^{Y}(1;0) | H_0 \sim N \left[ 0, \sum_{h = 1}^k \sigma^2_{\varepsilon} \left( \sum_{i = 0}^{k-h} b_i \sum_{j = i}^{k-h} \psi_{k-h-j}\right)^2 \right]
\end{equation*}

\begin{equation*}
    \hat{\bar{\tau}}_{t^*+k}^{Y}(1;0) | H_0 \sim N \left[ 0, \frac{\sigma^2_{\varepsilon}}{k^2} \sum_{h = 1}^k \left( \sum_{i = 0}^{k-h} b_i \sum_{j = i}^{k-h} \psi_{k-h-j}\right)^2 \right].
\end{equation*}

\clearpage

\end{appendices}

\end{document}